\documentclass[11pt, a4paper, oneside, reqno]{article}
\usepackage[total={7in,   10in}]{geometry}
\date{}
\usepackage{authblk} 
\usepackage[numbers,compress]{natbib}

\usepackage{amsmath,amssymb,mathrsfs,amsfonts} 
\usepackage{bm,bbm,dsfont} 

\usepackage{color,graphicx,graphics, epstopdf} 

\usepackage{algorithm,algorithmicx,multirow, titlecaps}  
\usepackage[english]{babel}
\usepackage[noend]{algpseudocode} 
\usepackage{threeparttable}
\usepackage{tikzsymbols}
\usepackage{multirow, titlecaps} 
\usepackage{tabularx}
\usepackage{booktabs}
\usepackage{multirow}
\usepackage{courier}
\usepackage{float}
\usepackage{multirow, titlecaps} 
\usepackage{tabularx}
\usepackage{booktabs}
\usepackage{multirow}
\usepackage{courier}
\usepackage{float}
\usepackage{makecell}
\usepackage{diagbox}
\usepackage{physics}
\usepackage{tikz}
\usepackage{circuitikz}

\allowdisplaybreaks
\usepackage{verbatim}

\DeclareMathAlphabet{\mathbbold}{U}{bbold}{m}{n}

\usepackage{xspace}

\usepackage{url}
 \iftrue    
    \usepackage[colorlinks=true]{hyperref}
	\usepackage{xcolor}
	\hypersetup{
		colorlinks,
		linkcolor={blue!90!black},
		citecolor={red!90!black},
		urlcolor={blue!20!black}
	}
	\allowdisplaybreaks
\fi

\newcommand{\mtiny}[1]{{\scalebox{.81}{#1}}}

\newcommand{\smtiny}[1]{{\scalebox{.5}{#1}}}
\newcommand{\stiny}[1]{{\scalebox{.38}{#1}}}

\renewcommand{\tr}{{\text{\mtiny{$\mathsf{T}$}}}}              
\iftrue

\else

\fi
\newcommand{\vc}[1]{{ \mathrm{#1} }}  
\newcommand{\mx}[1]{{ \mathrm{#1} }}  

\newcommand{\Scal}{{\mathcal{S}}}

\iffalse 
    \input{mystyleproof}
    \theoremstyle{plain}
    \newtheorem{theorem}{Theorem}
    \newtheorem{definition}{Definition}
    \newtheorem{assumption}{Assumption}
    \newtheorem{proposition}[theorem]{Proposition}
    \newtheorem{corollary}[theorem]{Corollary}
    \newtheorem{lemma}[theorem]{Lemma}
    \newtheorem{remark}{Remark}
    \newtheorem{example}{Example}
    \newtheorem{problem}{Problem}

\else

    \usepackage{amsthm}
    \renewcommand{\qedsymbol}{$\blacksquare$}

    \newtheorem{theorem}{Theorem}
    \newtheorem*{theorem*}{Theorem}
    
    \newtheorem*{definition*}{Definition}

    \newtheorem{corollary}[theorem]{Corollary}
    \newtheorem{lemma}[theorem]{Lemma}
    \newtheorem{remark}{Remark}
    
    \newtheorem*{example*}{Example}

    \newtheorem*{problem*}{Problem}
\fi

\newcommand{\mxA}{\mx{A}} 
\newcommand{\mxD}{\mx{D}} 
\newcommand{\mxM}{\mx{M}} 
\newcommand{\mxS}{\mx{S}} 

\newcommand{\vcu}{\vc{u}} 
\newcommand{\vcv}{\vc{v}} 
\newcommand{\vcw}{\vc{w}} 
\newcommand{\vcy}{\vc{y}} 

\newcommand{\Nu}{n_\vc{u}} 

\newcommand{\nsP}{{n_{\stiny{$\mathrm{P}$}}}} 
\newcommand{\nsR}{{n_{\stiny{$\mathrm{R}$}}}} 
\newcommand{\nP}{{n_{\smtiny{$\mathrm{P}$}}}} 
\newcommand{\nR}{{n_{\smtiny{$\mathrm{R}$}}}} 

\title{\LARGE \textbf{Safe Performance-aware Online Input Design}}
\author[$\dagger$]{Diyou Liu}
\author[$\star$]{Mohammad Khosravi}
\affil[$\dagger\star$]{{\small Delft Center for Systems and  Control, Delft University of Technology, Delft, Netherlands
\authorcr
{\texttt{
\{d.liu-9,mohammad.khosravi\}@tudelft.nl}}}}  
\begin{document}
\maketitle
\begin{abstract}
In this paper, we propose an input design framework for the safe online excitation and identification of uncertain linear systems without requiring informative initial data. The objective is to actively excite the system and collect sufficiently informative input-output data while maintaining safe operation and satisfactory control performance. 
Achieving this objective is challenging due to the potential conflict between the excitation required for learning and the safety and performance requirements, particularly when limited prior knowledge of the system is available.
Inspired by the dual control principle, the proposed scheme balances exploitation for control performance with exploration for model improvement. A Bayesian learning approach is employed to update the model uncertainty online, starting from a conservative initial uncertainty set. As informative data are collected, the uncertainty set contracts, enabling progressively less conservative operation. To reduce the online computational burden, the resulting program is reformulated into a computationally tractable form. Furthermore, an online ellipsoidal terminal set with non-increasing conservatism is designed to guarantee the recursive feasibility of the proposed scheme. Numerical experiments are conducted to demonstrate the effectiveness of the proposed framework, particularly in terms of uncertainty reduction and control performance.
\end{abstract}
    
\section{Introduction}\label{sec:Introduction}
System identification aims to obtain a model of an underlying system using measured input-output data. In practice, identifying a model that is uniformly accurate across all operating regions can be inefficient or unnecessary when the model is intended for a specific downstream task \cite{gevers2005identification, zadeh1962circuit}.
This limitation motivates objective-oriented identification, which prioritizes reducing uncertainty in the model components that most strongly influence the desired control objective \cite{anderson2024control}. Objective-oriented identification can be performed either offline  \cite{larsson2011optimal,annergren2017application}, using a historical dataset, or online  \cite{lee2024active,liang2020learning}, by recursively updating the model as new measurements become available. When online identification is conducted while simultaneously pursuing a control objective, identification and control become coupled because the applied inputs affect both closed-loop performance and the informativeness of the data collected for learning the system.
This motivates online input design, in which inputs are selected to reduce uncertainty in the regions most relevant to task performance while maintaining acceptable closed-loop behavior. This requirement is closely related to the dual control problem  \cite{aastrom1983theory,aastrom1995adaptive,feldbaum1960dual,milito1982innovations}, in which control inputs serve the dual purpose of regulating the system and exciting its uncertain dynamics. 
In addition to improving the informativeness of the collected data, practical online input design also needs to ensure safety throughout the identification process \cite{khosravi2022safety}. However, acquiring informative data from an unknown or highly uncertain system is particularly challenging under safety requirements \cite{rivera2003plant,wahlberg2006optimal}. When the underlying dynamics are not sufficiently understood, persistently exciting inputs may induce unexpected or undesirable system responses, potentially leading to constraint violations, unsafe transients, degraded reliability, or damage to system components \cite{narasimhan2011optimal}. These risks limit the extent to which aggressive exploration can be employed during identification. 

In recent years, a growing body of research has investigated safe online learning, input design, and dual control strategies \cite{khojasteh2020probabilistic,lew2022safe,dhiman2021control}. In \cite{zimmer2018safe}, an active learning framework based on Gaussian processes is proposed  to sequentially generate informative input trajectories while satisfying safety constraints. This method is initialized with previously collected safe trajectories and relies on prior knowledge to identify an initial safe region. In \cite{koller2018learning}, a learning-based MPC framework is considered  to enable safe exploration that provides probabilistic safety guarantees and incorporates terminal constraints to ensure recursive feasibility. 
This framework assumes prior knowledge of a safe terminal region in the state space and an associated control policy that renders the region positively invariant, which may be difficult to obtain in practice. 
In  \cite{ahmadi2021safely}, a framework is developed to sequentially learn unknown dynamical systems while ensuring that the system remains in the safe region. Rather than designing inputs during continuous closed-loop operation, the method depends on repeatedly stopping and resetting the system between short runs, with formal safety guarantees restricted to a few steps. 
Safe exploration for identifying unknown linear systems is considered in \cite{lu2017safe}, where safe action regions are gradually expanded around an initially known nominal safe action while model uncertainty is reduced. The approach assumes the availability of such an initial safe action and focuses on safe parameter identification rather than jointly optimizing exploration and closed-loop control performance.
More recently, \cite{hu2025closed} introduced a bounded probing signal into the closed-loop control input to provide exploration for system identification, while recursively designing this signal to keep the resulting output perturbation within prescribed limits. The method constrains only the incremental output deviation induced by the probing signal and assumes the availability of a known controller that renders the nominal closed-loop system exponentially stable. Thus, it does not explicitly guarantee satisfaction of general state or output safety constraints.
In \cite{de2026goal}, a goal-oriented safe active learning framework is integrated with MPC to collect informative data online while ensuring safety, recursive feasibility, finite-time termination of exploration, and satisfactory control performance. Nevertheless, this framework assumes that the internal dynamics of a pre-trained recurrent neural network are known, its hidden state is fully accessible, and the knowledge of a suitable safe invariant terminal set. Related adaptive and robust MPC approaches have also incorporated online model updates within constrained predictive control. For example, \cite{lorenzen2019robust} combines online set-membership identification with robust MPC to reduce conservatism while maintaining constraint satisfaction. Other adaptive MPC approaches address online parameter estimation with an emphasis on recursive feasibility and robust constraint satisfaction \cite{zhang2020adaptive,tanaskovic2014adaptive}. Within the dual control literature, \cite{parsi2022explicit} proposes an explicit dual MPC approach that balances reference tracking performance and active uncertainty reduction while guaranteeing robust constraint satisfaction and recursive feasibility. However, this method results in a nonlinear optimization problem that may be computationally demanding for online implementation. 
Due to the potential for improving control performance during online learning, one may also find data-driven control approaches \cite{tang2022data,coulson2019data} closely related to this line of research, which design controllers or control inputs directly from measured data without explicitly constructing a parametric model of the underlying dynamics \cite{markovsky2023data,dorfler2022bridging}. The performance of such data-driven control methods depends on whether the available data are sufficiently informative for the intended analysis or control task \cite{van2020data}. Moreover, ensuring safe data collection may require further developments in safety critical conditions, particularly when the dynamics of the system is highly complex and uncertain.

In summary, existing approaches focus on different aspects of safe online input design. Safe online learning methods often require prior knowledge of safe trajectories, control policies, or terminal sets, whereas adaptive and robust MPC methods primarily emphasize constraint satisfaction during model adaptation. Dual control formulations explicitly balance exploration and exploitation but may involve computationally demanding optimization. Consequently, simultaneously achieving active exploration, performance-aware exploitation, robust constraint satisfaction, recursive feasibility, and computational tractability without requiring informative initial data or prior knowledge of safe terminal sets and stabilizing controllers remains challenging.

To address the above challenges, this paper presents a framework for safely identifying unknown linear systems when no informative initial input-output data is available. A Bayesian approach is adopted to learn the unknown parameters, with prior model uncertainty represented by a conservative initial uncertainty set. To satisfy safety requirements, the learning process proceeds cautiously during the initial stage, resulting in relatively slow adaptation. As informative data are collected, the uncertainty set contracts, allowing progressively less conservative control and exploration. The control task is integrated into the problem, with a cost function that explicitly balances exploitation and exploration. For the exploitation component of the objective function, we consider two alternative formulations based on the worst-case and expected costs, respectively. A finite prediction horizon is employed to account for the future effects of the control inputs on both control performance and exploration. In addition, a suitable ellipsoidal online terminal set is designed to ensure recursive feasibility of the optimization problem. The resulting problem is reformulated as a tractable optimization problem. In the case of the worst-case cost, an additional approximation is introduced to obtain a convex formulation in terms of linear matrix inequalities (LMIs). Numerical studies are used to assess the quality of this approximation. Further simulations demonstrate the reduction of the uncertainty set and the resulting control performance. Finally, the proposed framework is extended to MIMO systems.
The main contributions are listed as follows:
\begin{itemize}
    \item Development of a safe online input-design framework for uncertain linear systems that does not require an informative initial dataset.
    \item Formulation of an online objective that explicitly balances exploration for uncertainty reduction and exploitation for control performance.
    \item Development of convex formulations and approximations to improve the computational tractability of the proposed online input design framework.
    \item Construction of an online ellipsoidal terminal set together with a proof of recursive feasibility under model uncertainty.
    \item Development of an active input design strategy with bounded random perturbations that guarantees asymptotic reduction of the parameter covariance.
    \item Evaluation of uncertainty reduction, approximation quality, and control performance through numerical studies.
\end{itemize}

The rest of this paper is structured as follows. 
In Section \ref{sec: Problem Formulation}, input design problem for performance-aware safe identification is formulated. 
In Section~\ref{sec: Technique Calculation}, we derive the optimization problem and present the proposed online performance-aware safety-constrained identification framework. In Section~\ref{sec:Convex Approximation}, a convex approximation of the problem with worst-case cost function is developed. In Section~\ref{sec:Extension to MIMO System}, we extend our proposed framework to MIMO system. In Section~\ref{sec:Actively Promoting Input Design}, we discuss actively promoting the input design to ensure the convergence of the eigenvalues of the parameter covariance matrix. Finally, in Section~\ref{sec: Numerical Experiments}, we show the numerical results to evaluate the proposed framework.

\textbf{\textit{Notation.}} 
Throughout this paper, $\mathbb{Z}$, $\mathbb{Z}_+$, $\mathbb{N}$ and $\mathbb{R}^{n\times m}$ are used to denote the set of integers, non-negative integers, positive integers and real numbers, and real-valued $n$ by $m$ matrices, respectively. Define $\mathbb{Z}_{m}$ and $\mathbb{N}_{m}$ as $\{0, 1, \ldots, m\}$ and $\{1, 2, \ldots, m\}$, respectively. Denote $q$ as the discrete time shift operator. The Euclidean norm of a vector $\mathrm{x} \in \mathbb{R}^n$ is denoted by $\| \mathrm{x} \|$. Moreover, the conditional probability of $A$ given $B$ is written as $p(A|B)$. Denote by $x_{k|t}$ the prediction of $x$ at time step $k$, obtained using the information available at time $t$.

\section{Input Design Problem for Performance-aware Online Safe Identification}
\label{sec: Problem Formulation}
Consider an \emph{unknown} stable LTI system $\Scal$ with impulse response $\mathrm{G}$. Our primary objective is to identify the system from the input-output measurement data. During the identification process, we additionally aim to achieve a desired control objective, e.g., by minimizing a prescribed control cost, while ensuring that the system outputs satisfy given safety constraints with high probability. Thus, we need to develop a suitable scheme for online input design that simultaneously facilitates accurate system identification, achieves satisfactory control performance, and maintains safe system operation.

\textbf{\textit{Mathematical Settings.}}
Since the system is stable, we know that its impulse response is absolute summable, and thus, it decays rapidly. Therefore, one can approximate the system by truncating 
$\mathrm{G}$
at a sufficiently large order $n_{\mathrm{g}} \in \mathbb{N}$.
Accordingly, the system can be represented by a finite impulse response (FIR) model as $(\mathrm{G}_k)_{k=0}^{n_{\mathrm{g}}-1}$.
Furthermore, let
$(\mathrm{u}_t)_{t \in \mathbb{Z}}$ and $(\mathrm{y}_t)_{t \in \mathbb{Z}}$ be the input and output of the system, respectively, where $\mathrm{u}_t \in \mathbb{R}^{n_{\mathrm{u}}}$ and $\mathrm{y} \in \mathbb{R}^{n_{\mathrm{y}}}$, for any $t \in \mathbb{Z}$. Suppose the output measurement is corrupted by noise, i.e., for any $t \in \mathbb{Z}$, we have
\begin{equation}\label{eq:y_t}
    \mathrm{y}_t = \sum_{k=0}^{n_\mathrm{g} - 1} \mathrm{G}_k^\tr \mathrm{u}_{t-k} + \mathrm{w}_t, 
\end{equation}
where $(\mathrm{w}_t)_{t \in \mathbb{Z}}$ is a sequence of independent and identically distributed (i.i.d.) Gaussian variables with
$\mathrm{w}_t \sim \mathcal{N}(0, \sigma_\mathrm{w}^2)$, for any $t \in \mathbb{Z}$.
Define matrix $\mu \in \mathbb{R}^{n_{\mathrm{g}} n_{\mathrm{u}} \times n_{\mathrm{y}}}$ as
\begin{equation}\label{eq:definition mu}
     \mu = 
     \big[
     \mathrm{G}_0^\tr,\,  \mathrm{G}_1^\tr,\, \ldots \, , \, \mathrm{G}_{n_\mathrm{g} - 1}^\tr
     \big]^\tr,
\end{equation}
collecting the truncated impulse response coefficients and vector $\varphi_t \in \mathbb{R}^{n_\mathrm{g} n_\mathrm{u}}$ as 
\begin{equation}\label{eq:definition phi}
    \varphi_t = 
    \big[
        \mathrm{u}_t^\tr,\,\mathrm{u}_{t-1}^\tr,\, \ldots \,,\, \mathrm{u}_{t-n_\mathrm{g} + 1}^\tr
    \big]^\tr,
\end{equation}
stacking the current and past $n_{\mathrm{g}} - 1$ input values.
Hence, following \eqref{eq:y_t}, $\mathrm{y}_t$ can be compactly written as
\begin{equation}
    \mathrm{y}_t = \mu^\tr \varphi_t + \mathrm{w}_t.
\end{equation}
Denote $\mu^i$ as the transpose of the i-th row of the matrix $\mu$. While the impulse response $\mu$ is assumed to be unknown, let \emph{prior information} about $\mu^i$ be provided as a Gaussian distribution, i.e., we have 
\begin{equation}\label{eq:prior information}
    \mu^i \sim \mathcal{N}({\mu}^i_0, {\Sigma}^i_0),    
\end{equation}
for any $ i \in \mathbb{N}_{n_\mathrm{y}}$,
where ${\mu}^i_0 \in \mathbb{R}^{n_\mathrm{g} n_\mathrm{u}}$ is a vector representing the mean of the prior distribution on parameter $\mu$, and ${\Sigma}^i_0 \in \mathbb{S}_{++}^{n_\mathrm{g} n_\mathrm{u} \times n_\mathrm{g} n_\mathrm{u} }$ is a positive definite matrix denoting its covariance and characterizing the uncertainty in $\mu^i$.

Given the prior information, while our aim is to execute the target control task and simultaneously update the estimates of the unknown parameter $\mu$, we need to satisfy an output safety constraint imposed on the system as
$\mu^\tr \varphi_t \in \mathcal{Y}_{\text{safe}}$,
for any $t$, where $\mathcal{Y}_{\text{safe}}$ is a set characterizing safety.
In practice, such constraints reflect safety requirements that must be respected during both control and parameter learning, thereby coupling performance objectives with safety considerations, which makes the online input design procedure challenging. 

\textbf{\textit{Problem Statement.}} Based on the above discussion, the main problem discussed in this paper is given below.

\noindent{\bfseries Problem}
\noindent{\normalfont\bfseries (Safe Performance-aware Online
Identification Problem)}\label{pro:problem formulation}
Given prior knowledge of the unknown parameter as in \eqref{eq:prior information} and a prescribled control objective, 
the aim is to design a control input to identify the impulse response of system in an online fashion, while performing the target control task and satisfying safety constraints.
\par\medskip

To address the introduced problem, we develop an optimization-based online input design scheme, taking into consideration the exploitation performance and safety maintenance.
More precisely, an approach based on Bayesian inference is employed to update the parameter estimates in a receding-horizon fashion as new measurement data become available. The resulting estimates are then used to construct suitable probabilistic robust parameter sets capturing the uncertainty in the system model, and incorporated into the problem to ensure safety with high probability during system operation. Thus, the online input design is formulated in a predictive fashion, solving an optimization problem with an objective function balancing exploitation and exploration, enabling the controller to maintain control performance while actively collecting informative data for identification. Further details are provided in the next section.

\section{Safe Performance-aware Online Input Design Scheme}\label{sec: Technique Calculation}
In this section, we introduce the structure of our optimization-based predictive online input design scheme. Since a primary concern for the practical implementation is the maintenance of safety, we first develop suitable nested robust parameter sets 
characterizing the uncertainty in $\mu$. To ensure, with high probability, that the true parameter lies in the intersection of all robust parameter sets, we construct these sets using a Bayesian linear regression, in line with \cite{abbasi2011improved} and \cite{dani2008stochastic}. Furthermore, to ensure the existence of a feasible solution at each time step during online implementation, the robust parameter sets are constructed such that the resulting optimization problem remains recursively feasible.
For ease of discussion, we first consider the single-output case, i.e., $n_{\mathrm{y}} = 1$. The extension to MIMO systems is discussed in Section~\ref{sec:Extension to MIMO System}.

\subsection{Online Input Design Scheme}\label{sec:Online Input Design Scheme}
To predict future system behavior and account for the long-term effects of the applied inputs, we adopt the receding-horizon strategy commonly used in MPC. At each time $t$, the control input is determined by solving a finite horizon optimization problem over a moving prediction window, but only the first element of the optimized input sequence is applied to the system. The prediction window is then shifted forward and the procedure is repeated using the latest measurements. To determine such an input sequence, an objective function is designed that evaluates both the predicted control performance and the informativeness of the resulting data. Consequently, one can comprises two components: exploitation and exploration. The exploitation component quantifies the performance with respect to the control task. 
More precisely, at time $t\in\mathbb{Z}_+$, we aim to minimize an exploitation cost 
$J_t(\mathbf{u}_{t,\nsP})$, 
where 
$\nP$ indicates the time horizon considered for the exploitation,
$\mathbf{u}_{t,\nsP} := (\vcu_{i|t})_{u=0}^\nsP$ denotes the related vector of the predicted inputs,
and 
$J_t:
\prod_{i=0}^{\nsP}\mathbb{R}^{\Nu}
\to\mathbb{R}$ indicate the corresponding control cost.
On the other hand, the exploration component promotes the acquisition of informative data and is denoted by $I_t(\mathbf{u}_{t,\nsR})$,
where $\nR$ 
denotes the receding horizon for exploration, 
$\mathbf{u}_{t,\nsR} := (\vcu_{i|t})_{u=0}^\nsR$ is the correspond vector of the predicted inputs,
and 
$I_t:
\prod_{i=0}^{\nsR} \mathbb{R}^{\Nu}
\to\mathbb{R}$ is an informativity index.

While the objective function determines the desired balance between exploitation and exploration, the optimized input sequence must also satisfy the operational constraints of the system. Accordingly, let $\mathcal{U}$ denote the set of admissible inputs. Then $\vcu_{k|t}$ needs to be in this admissible set for all prediction steps.
For output safety, we consider a probabilistic formulation.
Let $\mathrm{y}_{k|t} = \mu^\tr \varphi_{k|t}$ denote the predicted output. Output safety is imposed through a chance constraint as
\begin{equation}
    \mathds{P} \left( \mathrm{y}_{k|t} \in \mathcal{Y}_{\mathrm{safe}} \right) \geq 1- \delta_{\tau_t},
\end{equation}
where $\tau_t$ counts the number of updates of the parameter estimates up to time $t$, while $\delta_{\tau_t} \in (0,1)$ 
represents the allowable probability that the safety constraint is violated. Additionally, $\delta_{\tau_t}$ is chosen to decrease with $\tau_t$, thereby increasing the required probability of safety as the input design procedure progresses. With an appropriate summable choice of this sequence, a high probability safety guarantee can be established for the entire online procedure, as discussed in detail below. Furthermore, in Section~\ref{sec:Robust Safety Constraints}, we discussed how to formulate this constraint as LMIs, which can be efficiently solved online.


The above cost function and safety constraints depend on the parameter estimates and uncertainty.
Once the input is generated, it is applied to the system, and the resulting input-output data are collected and used to update these estimates and uncertainty.
Let $\mathcal{F}_t$ denote the $\sigma$-algebra generated by the history up to time $t$, defined as
\begin{equation} 
    \mathcal{F}_t := \sigma\big(\{\varphi_0, y_0, \ldots, \varphi_{t-1}, y_{t-1}\} \big), 
\end{equation}
where $\sigma(\cdot)$ denotes the $\sigma$-algebra generated by its argument.
Consider the posterior distribution of $\mu | \mathcal{F}_{t+1}$. Since noise $\mathrm{w}_t$ follows a  Gaussian distribution $\mathcal{N}(0, \sigma_\mathrm{w}^2)$, the posterior distribution also remains Gaussian, i.e., $\mu|\mathcal{F}_{t+1} ~\sim ( \tilde{\mu}_{t+1}, \tilde{\mathrm{A}}_{t+1})$, where $\tilde{\mu}_{t+1}$ and $ \tilde{\mathrm{A}}_{t+1}$ can be found by Bayesian law as
\begin{equation}\label{eq:bayesian linear regressor}
    \begin{split}
        \tilde{\mathrm{A}}_{t+1} &:= \tilde{\mathrm{A}}_{t} + \frac{1}{\sigma_\mathrm{w}^2} \varphi_{t} \varphi_{t}^\tr,\\
        \tilde{\mu}_{t+1} &:= \tilde{\Sigma}_{t+1} \Big( \tilde{\mathrm{A}}_{0} \tilde{\mu}_{0} + \frac{1}{\sigma_\mathrm{w}^2} \sum_{\tau<t+1} \varphi_\tau y_\tau  \Big),\\
        \tilde{\mathrm{A}}_t &:= \tilde{\Sigma}^{-1}_t,
    \end{split}
\end{equation}
for any $t$, with $\tilde{\Sigma}_0 := \Sigma_0$, and $\tilde{\mu}_0 := \mu_0$.
Based on the above posterior distribution, one needs to construct a sequence of robust parameter sets for safety and practical implementation considerations, which is discussed below in detail.

In summary, the online optimization problem can be formulated as
\begin{subequations}
\begin{align}
    &\min_{\{\mathrm{u}_{k|t}\}_{k=0}^{\nsP}} \; \, J_t(\mathbf{u}_{t,\nsP}) - \omega I_t(\mathbf{u}_{t,\nsR}) \\
    & \quad \ \, 
    \mathrm{s.t.} 
    \ \, \quad
    \mathrm{u}_{k|t} \in \mathcal{U}, \\
    & \qquad \qquad \; \,  
    \mathds{P} \left( \mathrm{y}_{k|t} \in \mathcal{Y}_{\mathrm{safe}} \right) \geq 1- \delta_{\tau_t}\\
    & \qquad \qquad \; \,  
    \varphi_{\nsP|t} \in \mathcal{R}_t,\\
    & \qquad \qquad \; \,  \forall k \in \mathbb{Z}_{\nsP-1},
\end{align}
\end{subequations}
where $\omega$ is a tuning weight that balances the exploitation and exploration term, and $\mathcal{R}_t$ denotes the online terminal set. As discussed, this online terminal set is designed to ensure the existence of a feasible solution throughout the input design procedure. The terminal set is updated online to reduce the conservatism because it depends on the parameter uncertainty, which evolves throughout the input design procedure. Details are provided in Section~\ref{sec:Termianl Set}.

\subsection{Robust Parameter Set Design}\label{sec:robust parameter set design}

Given that the true parameters follow the prior distribution $\mathcal{N}(\mu_0, \Sigma_0)$, we define set $\mathcal{C}_0$ as
\begin{equation}
    \mathcal{C}_0 := \{\nu : \|\nu - \mu_0\|_{\Sigma_0^{-1}} \leq \beta_0^{\frac{1}{2}} \},    
\end{equation}
where $\beta_0 = \chi^2_{n_{\mathrm{g}}} (1 - \delta_0)$, and $\delta_0$ is a positive parameter close to 0. This set characterizes the initial uncertainty in the parameter vector.
Subsequently, we need to design a suitable sequence of robust parameter sets $(\mathcal{C}_t)_{t \in \mathbb{N}}$ such that the nested condition, i.e., $\mathcal{C}_{t+1} \subseteq \mathcal{C}_t$,
holds for any $t \in \mathbb{Z}_+$, where $\mathcal{C}_t$ is an ellipsoid with structure 
\begin{equation}
\begin{split}
    \mathcal{C}_t &:= \{\nu : \|\nu - \mu_t\|_{\Sigma_t^{-1}} \leq \beta_t^{\frac{1}{2}} \},
\end{split}
\end{equation}
with $\mxA_t$ being defined as $\mathrm{A}_t := \Sigma_t^{-1}$.
The role of the above nested condition is to ensure the recursive feasibility of the optimization program.

In addition to the nested condition, we also require the probability condition, i.e., $\mu$ is in the intersection of all robust parameter set $(\mathcal{C}_t)_{t\in\mathbb{Z}_+}$ with a high probability larger than $1-\delta$, where $\delta$ is a positive parameter representing the maximal probability that the true parameter is not contained by all robust parameter sets.
This probability condition ensures that the subsequent safety constraints are constructed based on a valid uncertainty set. 

Define posterior credible set as
\begin{equation}
    \tilde{\mathcal{C}}_t (\tilde{\beta}_t) 
    := 
    \{\nu \mid \|\nu - \tilde{\mu}_t\|_{\tilde{\Sigma}_t^{-1}} \leq \tilde{\beta}_t^{\frac{1}{2}} \}.
\end{equation} 
A natural candidate for $\mathcal{C}_t$ is $\tilde{\mathcal{C}}_t(\tilde{\beta}_t)$, where $\tilde{\beta}_{t}$ needs to be suitably chosen to make both nested condition and probability condition satisfied. However, finding such $\tilde{\beta}_{t}$ is challenging, as a large $\tilde{\beta}_{t}$ is needed for the probability condition and a small $\tilde{\beta}_{t}$ is needed for the nested condition. To balance these competing requirements, we do not update the robust parameter set at every time step. Instead, the update times are selected such that both conditions are satisfied.
The robust parameter set is constructed according to the following procedure.

Consider a certain step $t \in \mathbb{N}$. Let $\tau_t$ denote the update index, defined as one plus the number of completed update steps before time $t$. Define the lower bound $\tilde{\beta}_t^{\min}$ as
\begin{equation}
    \tilde{\beta}_t^{\min}  := \chi^2_{n_{\mathrm{g}}}(1 - \delta_{\tau_t}),
\end{equation}
 where $\delta_{\tau_t} \in (0,1)$, and $\tilde{\beta}_{0}^{\min} := \beta_0$.
Additionally, define the upper bound $\tilde{\beta}_t^{\max}$ via the optimization problem
\begin{equation}\label{eq:biggest nesting set}
    \begin{split}
        \tilde{\beta}_{t}^{\max} := 
        &
        \max_{\beta \geq 0, \lambda \geq 0} \    \beta \\
        & \quad 
        \ \mathrm{s.t.}
        \quad 
    \begin{bmatrix}
        \tilde{\mathrm{A}}_{t} \!-\! \lambda  \mathrm{A}_{t-1} & -\tilde{d}_{t} \!+\! \lambda d_{t-1}  \\
        \star & \tilde{a}_{t} \!-\! \beta \!-\! \lambda a_{t-1} \!+\! \lambda \beta_{t-1}
    \end{bmatrix}
        \succeq 0,
    \end{split}
\end{equation} 
where $d_t = \mathrm{A}_{t} \mu_t$, $\tilde{d}_t = \tilde{\mathrm{A}}_{t} \tilde{\mu}_{t}$, $a_t = \mu_{t}^\tr \mathrm{A}_{t}\mu_{t}$, and $\tilde{a}_t = \tilde{\mu}_{t}^\tr \tilde{\mathrm{A}}_{t}\tilde{\mu}_{t}$, initialized with $\tilde{\beta}_{0}^{\max} := \beta_0$.

Let $t_0 = 0$, and $t_1, t_2, \ldots$ denote the sequence of time instants at which the interval $[\tilde{\beta}_{t_i}^{\min}, \tilde{\beta}_{t_i}^{\max}]$ is a non-empty set. Define the time index set $\mathcal{T}_i$, for $i \in \mathbb{Z}_+$, as 
\begin{equation}
    \mathcal{T}_i := \{t \in \mathbb{Z}_+ | t_i \leq t < t_{i+1}\}.
\end{equation}
Subsequently, the robust parameter set $\mathcal{C}_t$ is defined according to the following rules as
\begin{equation}\label{eq:robust parameter set update}
\begin{split}
    \mathcal{C}_{t} &:= \tilde{\mathcal{C}}_{t_i}(\tilde{\beta}_{t_i}^{\max}),
\end{split}
\end{equation}
for any $ t \in \mathcal{T}_i$. The procedure for constructing the robust parameter set is summarized in Algorithm~\ref{Al:Update Robust Parameter Set}.


\begin{algorithm}[t]
    \caption{Update Robust Parameter Set}\label{Al:Update Robust Parameter Set}
    \begin{algorithmic}
        \Statex \textbf{Input:} $\mathrm{A}_{t-1}, \mu_{t-1}, \beta_{t-1}, \tilde{\mathrm{A}}_{t-1}, \tilde{\mu}_{t-1}, \mathrm{y}_{t-1}, \varphi_{t-1}, i$.
        \Statex \textbf{Output:} $\mathrm{A}_{t}, \mu_{t}, \beta_{t}, i$.
            \State Compute posterior distribution $\mathcal{N}(\tilde{\mu}_{t}, \tilde{\Sigma}_{t})$ using \eqref{eq:bayesian linear regressor}
            \State Find $\tilde{\beta}_{t}^{\max}$ by solving \eqref{eq:biggest nesting set}
            \State Find $\tilde{\beta}_{t}^{\min} = \chi^2_{n_{\mathrm{g}}} (1 - \delta_{i})$
            \If{$\tilde{\beta}_{t}^{\max} \geq \tilde{\beta}_{t}^{\min}$}
            \State $\beta_{t} \gets \tilde{\beta}_{t}^{\max}$, $\mathrm{A}_{t} \gets \tilde{\mathrm{A}}_{t}$, $\mu_{t} \gets \tilde{\mu}_{t}$, and $i \gets i+1$
            \Else
            \State $\beta_{t} \gets \beta_{t-1}$, $\mathrm{A}_{t} \gets \mathrm{A}_{t-1}$, and $\mu_{t} \gets \mu_{t-1}$
            \EndIf
    \end{algorithmic}
\end{algorithm}

\begin{theorem}\label{thm:probability constraint}
Let prior knowledge $\mu \sim \mathcal{N}(\tilde{\mu}_0, \tilde{\Sigma}_0)$ be given, and $\beta_0 = \chi^2_{n_{\mathrm{g}}} (1 - \delta_0)$. Given an input sequences $(\mathrm{u}_t)_{t \in \mathbb{Z}}$, let $(\delta_i)_{i \in \mathbb{Z}_+}$ be a sequence such that $\sum_{i=0}^\infty \delta_i \leq \delta$.
Let the robust parameter set $\mathcal{C}_t$ be updated according to the rule in \eqref{eq:robust parameter set update}. Then, the nested condition and the probability condition hold simultaneously, i.e., we have
\begin{equation}
        \mathcal{C}_{t+1} \subseteq \mathcal{C}_{t}, \label{eq:parameter update constraints 1}
\end{equation}
for any $t \in \mathbb{Z}_+$, and
\begin{equation}
    \mathds{P}(\mu \in \bigcap_{\tau=0}^\infty \mathcal{C}_\tau) \geq 1 - \delta. \label{eq:parameter update constraints 2}
\end{equation}
\end{theorem}
\begin{proof}
    First, consider the nested condition in \eqref{eq:parameter update constraints 1}. Since $\mxA_t \succeq \mxA_0 \succ 0$, $\mxA_t$ is positive definite. Moreover, $\tilde{\beta}_t^{\min}>0$ for all $t\in\mathbb{Z}_+$, which implies that $\beta_t>0$ for all $t\in\mathbb{Z}_+$. Thus, $\mathcal{C}_t$ is a nonempty ellipsoid with a nonempty interior for all $t\in\mathbb{Z}_+$. If $[\tilde{\beta}_{t+1}^{\min}, \tilde{\beta}_{t+1}^{\max}]$ is a non-empty set, using S-procedure, the nested condition is equivalent to
    \begin{equation}
            \begin{bmatrix}
        \tilde{\mathrm{A}}_{t+1} \!-\! \lambda  \mathrm{A}_{t} & -\tilde{d}_{t+1} \!+\! \lambda d_t  \\
        \star & \tilde{a}_{t+1} \!-\! \beta \!-\! \lambda a_t \!+\! \lambda \beta_t
    \end{bmatrix}
        \succeq 0.
    \end{equation}
    Thus, program in \eqref{eq:biggest nesting set} finds the maximal $\tilde{\beta}_{t+1}$ such that $\tilde{\mathcal{C}}_{t+1}(\tilde{\beta}_{t+1}) \subseteq \mathcal{C}_t$.
    According to the update rule \eqref{eq:robust parameter set update}, we have $\mathcal{C}_{t+1} = \tilde{\mathcal{C}}_{t+1}(\tilde{\beta}_{t+1}^{\max}) \subseteq \mathcal{C}_t$. If $[\tilde{\beta}_{t+1}^{\min}, \tilde{\beta}_{t+1}^{\max}]$ is an empty set, $\mathcal{C}_{t+1}$ is the set same as $\mathcal{C}_t$, which also implies that $\mathcal{C}_{t+1} \subseteq \mathcal{C}_{t}$. Therefore, the nested condition holds.
    
    Regarding the probability condition, for time step $t$ in the interval $t_i \leq t < t_{i+1}$, the probability that $\mu \notin \mathcal{C}_t$ is the same as the probability that $\mu \notin \mathcal{C}_{t_i}$. Thus, we have
    \begin{equation}
    \begin{split}
        \mathds{P}(\, \exists t, \ \mu \notin \mathcal{C}_t) 
        &= \mathds{P}(\, \exists i, \ \mu \notin \mathcal{C}_{t_i}),\\
        &\leq 
        \sum_{i=0}^\infty \mathds{P} \left(\mu \notin \mathcal{C}_{t_i} \right),\\
    \end{split}
    \end{equation}
    where the inequality is due to the union bound. Using the law of total expectation, this equation can be written as
    \begin{equation}
        \begin{split}
            \sum_{i=0}^\infty \mathds{P} \!\left(\mu \notin \mathcal{C}_{t_i} \right)
            =& \sum_{i=0}^\infty \mathds{E}\left[\mathds{P}\left(\mu \notin \mathcal{C}_{t_i} | \mathcal{F}_{t_i}\right)\right].
        \end{split}
    \end{equation}
    Substituting the definition of the ellipsoid $\mathcal{C}_{t_i}$, we have
    \begin{equation}
        \begin{split}
            \mathds{P}\left(\mu \notin \mathcal{C}_{t_i} | \mathcal{F}_{t_i}\right) &= \mathds{P} \left( \|\mu - \tilde{\mu}_{t_i}\|^2_{\tilde{\mathrm{A}}_{t_i}} \!>\! \tilde{\beta}_{t_i}^{\max} | \mathcal{F}_{t_i} \right)\\
            &\leq \mathds{P} \left( \|\mu - \tilde{\mu}_{t_i}\|^2_{\tilde{\mathrm{A}}_{t_i}} \!>\! \tilde{\beta}_{t_i}^{\min} | \mathcal{F}_{t_i} \right)\\
            &= \mathds{P} \left( \|\mu - \tilde{\mu}_{t_i}\|^2_{\tilde{\mathrm{A}}_{t_i}} \!>\! \chi_{n_\mathrm{g}}^2(1-\delta_i) | \mathcal{F}_{t_i} \right).
        \end{split}
    \end{equation}
    Moreover, since $\mu | \mathcal{F}_{t_i} \sim \mathcal{N}(\tilde{\mu}_{t_i}, \tilde{\Sigma}_{t_i})$, we have
    \begin{equation}
        \|\mu - \mu_{t_i}\|^2_{\mathrm{A}_{t_i}} | \mathcal{F}_{t_i} \sim \chi_{n_\mathrm{g}}^2,
    \end{equation}
    which implies that
    \begin{equation}
    \begin{split}
         \mathds{P} \left( \|\mu - \mu_{t_i}\|^2_{\mathrm{A}_{t_i}} \!>\! \chi_{n_\mathrm{g}}^2(1-\delta_i) | \mathcal{F}_{t_i} \right) = \delta_{i}.
    \end{split}
    \end{equation}
    Thus, we have
    \begin{equation}
        \begin{split}
            \sum_{i=0}^\infty \mathds{E}\left[\mathds{P}\left(\mu \notin \mathcal{C}_{t_i} | \mathcal{F}_{t_i}\right)\right]
            \leq& \sum_{i=0}^\infty \mathds{E}\left[\delta_i\right]
            \leq \delta.
        \end{split}
    \end{equation}
    This concludes the proof.
\end{proof}
\begin{remark}
\normalfont
 The Algorithm~\ref{Al:Update Robust Parameter Set} guarantees that, for an infinite-horizon control task, the true parameter belongs to the intersection of the robust parameter sets with high probability. In practice, if one is given a task to track a finite reference with length $T$, it is sufficient to use a constant $\delta$ instead of reducing it after every update, so that the true parameter is covered by the intersection of $\mathcal{C}_t$, for all $t \in \mathbb{Z}_T$, with a high probability.
\end{remark}


\subsection{Robust Safety Constraints}\label{sec:Robust Safety Constraints}
The primary concern addressed in the paper is system safety. In real-world applications, it is common for practical systems to impose constraints in order to protect the system from unsafe operating conditions. Such constraints ensure that the system remains within safe limits, preventing potential damage to components, avoiding unsafe behaviors, and maintaining overall reliability. 

Consider the safety constraint as
\begin{equation}\label{eq:safety chance constraint}
    |\mu^\tr \varphi_t| \leq \bar{\mathrm{y}},  
\end{equation}
for all $t \in \mathbb{Z}_+$.  
Since $\mu$ is unknown, we need to ensure robust safety despite the presence of model uncertainties within the prediction horizon. From the update law \eqref{eq:robust parameter set update}, we have $\mathds{P}(\mu \in \mathcal{C}_t) \geq 1 - \delta_{\tau_t}$. To establish safety guarantees under this probabilistic condition, the robust safety constraint needs to satisfy
\begin{equation}\label{eq:safety constraint}
    |\nu^\tr \varphi_{k|t}| \leq \bar{\mathrm{y}}, \quad \forall \nu \in \mathcal{C}_t,
\end{equation}
for all $k \in \mathbb{Z}_{\nsP}$. The above constraint is not tractable, since it needs to be satisfied for all $\nu \in \mathcal{C}_t$. Thus, in practice, to achieve a tractable and convex formulation, we apply the following theorem.

\begin{theorem}\label{thm:safety}
    The safety constraint \eqref{eq:safety constraint} is equivalently satisfied if and only if there exists symmetric matrices $\bar{\mathrm{D}}_{k|t}$ and scalars $\bar{\lambda}_{k|t}$ satisfying
    \begin{equation}\label{eq:safety constraints convex LMI}
    \begin{split}
        &\begin{bmatrix}
            \bar{\mathrm{D}}_{k|t} &  -\bar{\lambda}_{k|t} d_t\\
            \star & \bar{\lambda}_{k|t} a_t \!-\! \bar{\lambda}_{k|t} \beta_t \!+\! \bar{\mathrm{y}}^2 
        \end{bmatrix}
        \succeq 0, \\
        &\begin{bmatrix}
            \bar{\lambda}_{k|t}\mathrm{A}_t \!-\! \bar{\mathrm{D}}_{k|t} & \varphi_{k|t}\\
            \star & 1
        \end{bmatrix}
        \succeq 0,\\
        &\ \bar{\lambda}_{k|t} \geq 0,
    \end{split}
    \end{equation}
    for all $k \in \mathbb{Z}_{\nsP}$.
\end{theorem}
\begin{proof}
    First, the S-procedure is applied to express constraint \eqref{eq:safety constraint} in the following equivalent form:
\begin{equation}\label{eq:safety constraint S-prodecure}
\begin{split}
    &\begin{bmatrix}
        \bar{\lambda}_{k|t} \mathrm{A}_t \!-\! \varphi_{k|t} \varphi_{k|t}^\tr &  -\bar{\lambda}_{k|t} d_t\\
        \star & \bar{\lambda}_{k|t} a_t \!-\! \bar{\lambda}_{k|t} \beta_t \!+\! \bar{\mathrm{y}}^2 
    \end{bmatrix}
    \succeq 0,
    \\
    &\ \bar{\lambda}_{k|t} \geq 0,
\end{split} 
\end{equation}
for all $k \in \mathbb{Z}_\nsP$.
By introducing the auxiliary symmetric matrix $\bar{\mathrm{D}}_{k|t}$ to bound the non-convex term, constraint \eqref{eq:safety constraint S-prodecure} can be rewritten as,
\begin{subequations}\label{eq:safety constraints convex LMI tmp}
    \begin{align}
        &\begin{bmatrix}
            \bar{\mathrm{D}}_{k|t} &  -\bar{\lambda}_{k|t} d_t\\
            \star & \bar{\lambda}_{k|t} a_t \!-\! \bar{\lambda}_{k|t} \beta_t \!+\! \bar{\mathrm{y}}^2 
        \end{bmatrix}
        \succeq 0, \\
        &\ \bar{\lambda}_{k|t}\mathrm{A}_t - \varphi_{k|t}\varphi_{k|t}^\tr 
        \succeq \bar{\mathrm{D}}_{k|t} \\
        &\ \bar{\lambda}_{k|t} \geq 0.
    \end{align}
\end{subequations}
Using the Schur complement lemma, the above constraint is equivalent to
\begin{equation}
    \begin{split}
        &\begin{bmatrix}
            \bar{\mathrm{D}}_{k|t} &  -\bar{\lambda}_{k|t} d_t\\
            \star & \bar{\lambda}_{k|t} a_t \!-\! \bar{\lambda}_{k|t} \beta_t \!+\! \bar{\mathrm{y}}^2 
        \end{bmatrix}
        \succeq 0, \\
        &\begin{bmatrix}
            \bar{\lambda}_{k|t}\mathrm{A}_t \!-\! \bar{\mathrm{D}}_{k|t} & \varphi_{k|t}\\
            \star & 1
        \end{bmatrix}
        \succeq 0,\\
        &\bar{\lambda}_{k|t} \geq 0,
    \end{split}
\end{equation}
for all $k \in \mathbb{Z}_{\nsP}$. This completes the proof.
\end{proof}
Formulation \eqref{eq:safety constraints convex LMI} provides LMIs representation of the robust safety constraints. This formulation enables the constraints to be handled efficiently within the online predictive optimization problem using standard convex optimization tools, while preserving the high-probability safety guarantee.

\subsection{Terminal Set Design}\label{sec:Termianl Set}
As discussed previously, the online implementation requires the optimization problem to remain feasible at every time step. Thus, an online terminal set is designed and incorporated into the predictive optimization problem to guarantee this.
Define a projection matrix $\mathrm{P}$ as, 
\begin{equation}
    \mathrm{P} := \big[
        \mathbf{I}_{(n_\mathrm{g}-1)n_{\mathrm{u}}} \quad  0_{(n_{\mathrm{g}}-1) n_{\mathrm{u}} \times n_{\mathrm{u}}}
    \big].
\end{equation}
Using this notation, $\varphi_{t+1}$ can be written as $\varphi_{t+1} = \big[
    \mathrm{u}_{t+1}^\tr \  (\mathrm{P} \varphi_{t})^\tr
\big]^\tr$. Additionally, define a terminal control law $\mathrm{u}_{t+1} = \mathrm{K} \varphi_{t}$, where $\mathrm{K}$ is pre-computed such that $\mathrm{M} = \big[
    \mathrm{K}^\tr \  \mathrm{P}^\tr
\big]^\tr$ is a stable matrix.

Let the terminal set be parameterized as $\mathcal{R}_t := \{\varphi| \varphi^\tr \mathrm{S}_t \varphi \leq 1\}$, where $\mxS_t$ is a positive definite matrix. For each time step $t$, the objective is to construct a terminal set with non-increasing conservatism such that for any $\varphi_{t} \in \mathcal{R}_{t}$, we have
\begin{subequations}\label{eq:terminal constraints}
    \begin{align}
        \varphi_{t+1}^\tr \mathrm{S}_t \varphi_{t+1} &\leq \varphi_{t}^\tr \mathrm{S}_t \varphi_{t}, \label{eq:terminal constraints 1}\\
        \|\mathrm{K} \varphi_t\| &\leq \bar{\mathrm{u}}, \label{eq:terminal constraints 2}\\
        |\nu^\tr \varphi_{t}| &\leq \bar{\mathrm{y}}, \  \forall \nu \in \mathcal{C}_t. \label{eq:terminal constraints 3}
    \end{align}
\end{subequations}
Constraint \eqref{eq:terminal constraints 1} ensures that $\varphi_{t+1}$ remains in the terminal set, while \eqref{eq:terminal constraints 2} and \eqref{eq:terminal constraints 3} enforce the input and output safety constraints, respectively, for all admissible parameters.

The terminal set is updated online because constraint \eqref{eq:terminal constraints 3} depends on $\mathcal{C}_t$. Since $\mathcal{C}_t \subseteq \mathcal{C}_{t-1}$, constraint \eqref{eq:terminal constraints 3} becomes no more restrictive as the robust parameter set is updated, allowing the terminal set conservatism to be non-increasing. 

Characterizing such a terminal set is challenging because the underlying constraints are computationally intractable. In particular, the last constraint \eqref{eq:terminal constraints 3} must hold for all $\varphi_t \in \mathcal{R}_t$ and $\nu \in \mathcal{C}_t$, making it a robust constraint that must be satisfied for all admissible realizations. Thus, to mitigate online computational demands, the following convex optimization problem is formulated to explicitly determine $\mathcal{R}_{t}$.

\begin{theorem}\label{the:terminal constraints LMI}
    Let $\mathrm{K}$ be a control policy such that $\mathrm{M} = \big[ \mathrm{K}^\tr \  \mathrm{P}^\tr \big]^\tr$ is a stable matrix. Then constraints \eqref{eq:terminal constraints} hold for all $\varphi_t \in \mathcal{R}_t$ if and only if there exists non-negative values $\lambda_{T,1|t}, \lambda_{T,2|t}, \lambda_{T,3|t}$ such that we have
    \begin{subequations}\label{eq:terminal constraints LMI}
        \begin{align}
            & 
            \begin{bmatrix}
                \mathrm{S}_t & \mathrm{M}^\tr\mathrm{S}_t\\
                \mathrm{S}_t\mathrm{M} & \mathrm{S}_t
            \end{bmatrix} 
            \succeq 0, \label{eq:terminal constraints LMI 1}\\
            &
            \begin{bmatrix}
                \mathrm{S}_t & \mu_t \\ \mu_t^\tr & \bar{\mathrm{y}}^2
            \end{bmatrix}
            \succeq 0,\label{eq:terminal constraints LMI 3}\\
            &-\!\lambda_{T,1|t}\!
            \begin{bmatrix}
                \mathrm{K}^\tr\mathrm{K} & 0 \\ 0 & -\bar{\mathrm{u}}^2    
            \end{bmatrix}
            \!+\!
            \begin{bmatrix}
                \mathrm{S}_t & 0 \\ 0 & -1
            \end{bmatrix}
            \succeq 0, \label{eq:terminal constraints LMI 2}\\
            &-\!\lambda_{T,2|t}\!
            \begin{bmatrix}
                \bar{\mathrm{M}}_t & \bar{\mathrm{y}}  \mu_t  \\
                \star & -\bar{\mathrm{y}}^2
            \end{bmatrix} 
            \!+\!
            \begin{bmatrix}
                \mathrm{S}_t & 0 \\ 0 & -1
            \end{bmatrix}
            \succeq 0, \label{eq:terminal constraints LMI 4}\\
            &-\!\lambda_{T,3|t}\!
            \begin{bmatrix}
                \bar{\mathrm{M}}_t & -\bar{\mathrm{y}}  \mu_t  \\
             \star & -\bar{\mathrm{y}}^2
            \end{bmatrix} 
            \!+\!
            \begin{bmatrix}
                \mathrm{S}_t & 0 \\ 0 & -1
            \end{bmatrix}
            \succeq 0, \label{eq:terminal constraints LMI 5}
        \end{align}
    \end{subequations}
where $\bar{\mathrm{M}}_t = \beta_t \mathrm{A}_t^{-1} \!-\!\mu_t \mu_t^\tr $.
\end{theorem}
\begin{proof}
We start with the first constraint in \eqref{eq:terminal constraints 1}. Under the control law,  \eqref{eq:terminal constraints 1} can be written as
\begin{equation}
    \begin{split}
        \varphi_{t}^\tr \mathrm{M}^\tr
        \mathrm{S}_t
        \mathrm{M}\varphi_{t}
        \leq \varphi_{t} \mathrm{S}_t \varphi_t,
        \quad \forall \varphi_{t} \in \mathcal{R}_{t}.
    \end{split}
\end{equation}
Using Schur complement, the above equation can be reformulated to
\begin{equation}
    \begin{bmatrix}
        \mathrm{S}_t & \mathrm{M}^\tr\mathrm{S}_t\\
        \mathrm{S}_t\mathrm{M} & \mathrm{S}_t
    \end{bmatrix} \succeq 0.
\end{equation}
The second constraint \eqref{eq:terminal constraints 2} is 
\begin{equation}
    \varphi_t^\tr\mathrm{K}^\tr\mathrm{K}\varphi_t \leq \bar{\mathrm{u}}^2, \forall \varphi_t \in \mathcal{R}_t.
\end{equation}
Using S-procedure, it is equivalent to
\begin{equation}
    -\lambda_{T,1|t}
    \begin{bmatrix}
        \mathrm{K}^\tr\mathrm{K} & 0 \\ 0 & -\bar{\mathrm{u}}^2    
    \end{bmatrix}
    +
    \begin{bmatrix}
        \mathrm{S}_t & 0 \\ 0 & -1
    \end{bmatrix}
    \succeq 0.
\end{equation}
For the last constraint \eqref{eq:terminal constraints 3}, using Lemma~\ref{lem:|c^T a|}, it can be written as
\begin{subequations}\label{eq:y safety terminal constraint}
\begin{align}
    \left( \beta_t \varphi_t^\tr\mathrm{\mathrm{A}}_t^{-1} \varphi_t \right)^{\frac{1}{2}} + {\mathrm{\mu}}_t^\tr \mathrm{\varphi_{t}} &\leq \bar{\mathrm{y}}, \forall \varphi_{t} \in \mathcal{R}_{t}, \label{eq:y safety terminal constraint 1}\\
    \left( \beta_t \varphi_t^\tr\mathrm{\mathrm{A}}_t^{-1} \varphi_t \right)^{\frac{1}{2}} - {\mathrm{\mu}}_t^\tr \mathrm{\varphi_{t}} &\leq \bar{\mathrm{y}}, \forall \varphi_{t} \in \mathcal{R}_{t}.  \label{eq:y safety terminal constraint 2}
\end{align}
\end{subequations}
Constraint \eqref{eq:y safety terminal constraint 1} is equivalent to
\begin{subequations}\label{eq:y safety terminal constraint }
\begin{align}
    \bar{\mathrm{y}} - {\mathrm{\mu}}_t^\tr \mathrm{\varphi_{t}} &\geq 0, \label{eq:y safety terminal constraint 3}\\
    \left(\bar{\mathrm{y}} - {\mathrm{\mu}}_t^\tr \mathrm{\varphi_{t}} \right)^2 &\geq \beta_t \varphi_t^\tr\mathrm{\mathrm{A}}_t^{-1} \varphi_t, \label{eq:y safety terminal constraint 4}
\end{align}
\end{subequations}
holds $\forall \varphi_{t} \in \mathcal{R}_{t}$.
Applying Lemma~\ref{lem:|c^T a|} to \eqref{eq:y safety terminal constraint 3} and S-procedure \eqref{eq:y safety terminal constraint 4}, the above equations hold if and only if there exists $\lambda_{T,2|t} \geq 0$ such that
\begin{subequations}
    \begin{align}
        &\begin{bmatrix}
            \mathrm{S}_t & \mu_t \\ \mu_t^\tr & \bar{\mathrm{y}}^2
        \end{bmatrix}
        \succeq 0,\\
        &-\lambda_{T,2|t}\!
            \begin{bmatrix}
                \bar{\mathrm{M}}_t & \bar{\mathrm{y}} \mu_t  \\
             \star & -\bar{\mathrm{y}}^2
            \end{bmatrix}
            \!+\!
            \begin{bmatrix}
                \mathrm{S}_t & 0 \\ 0 & -1
            \end{bmatrix}
            \succeq 0.
    \end{align}
\end{subequations}
Similarly, \eqref{eq:y safety terminal constraint 2} is equivalent to the existence of $\lambda_{T,3|t} \geq 0$ such that
\begin{subequations}
    \begin{align}
        &\begin{bmatrix}
            \mathrm{S}_t & \mu_t \\ \mu_t^\tr & \bar{\mathrm{y}}^2
        \end{bmatrix}
        \succeq 0,\\
        &-\lambda_{T,3|t} 
            \begin{bmatrix}
                \bar{\mathrm{M}}_t & -\bar{\mathrm{y}}  \mu_t  \\
             \star & -\bar{\mathrm{y}}^2
            \end{bmatrix} 
            \!+\!
            \begin{bmatrix}
                \mathrm{S}_t & 0 \\ 0 & -1
            \end{bmatrix}
            \succeq 0.
    \end{align}
\end{subequations}
Combining all the constraints concludes the proof.
\end{proof}
Theorem \ref{the:terminal constraints LMI} shows that the constraints \eqref{eq:terminal constraints} can be formulated as LMIs, which facilitates online computation. To construct the terminal set, we minimize  $\mathrm{tr} (\mathrm{S}_t)$ subject to constraints \eqref{eq:terminal constraints LMI}. 
The optimization program is summarized as
    \begin{subequations}\label{eq:terminal constraints OP}
        \begin{align}
         &\min_{\mathrm{S} \succeq \epsilon \mathbf{I}}  \qquad
         \mathrm{tr}(\mathrm{S})\\
            & \mathrm{s.t.} \qquad
            \begin{bmatrix}
                \mathrm{S} & \mathrm{M}^\tr\mathrm{S}\\
                \mathrm{S}\mathrm{M} & \mathrm{S}
            \end{bmatrix} 
            \succeq 0, \label{eq:terminal constraints OP 1}\\
            & {\color{white} \mathrm{s.t.}} \qquad
            \begin{bmatrix}
                \mathrm{S} & \mu_t \\ \mu_t^\tr & \bar{\mathrm{y}}^2
            \end{bmatrix}
            \succeq 0,\label{eq:terminal constraints OP 3}\\
            & {\color{white} \mathrm{s.t.}} \qquad
            -\!\lambda_{T,1|t}\!
            \begin{bmatrix}
                \mathrm{K}^\tr\mathrm{K} & 0 \\ 0 & -\bar{\mathrm{u}}^2    
            \end{bmatrix}
            \!+\!
            \begin{bmatrix}
                \mathrm{S} & 0 \\ 0 & -1
            \end{bmatrix}
            \succeq 0, \label{eq:terminal constraints OP 2}\\
            & {\color{white} \mathrm{s.t.}} \qquad
            -\!\lambda_{T,2|t}\!
            \begin{bmatrix}
                \bar{\mathrm{M}}_t & \bar{\mathrm{y}}  \mu_t  \\
                \star & -\bar{\mathrm{y}}^2
            \end{bmatrix} 
            \!+\!
            \begin{bmatrix}
                \mathrm{S} & 0 \\ 0 & -1
            \end{bmatrix}
            \succeq 0, \label{eq:terminal constraints OP 4}\\
            & {\color{white} \mathrm{s.t.}} \qquad
            -\!\lambda_{T,3|t}\!
            \begin{bmatrix}
                \bar{\mathrm{M}}_t & -\bar{\mathrm{y}}  \mu_t  \\
             \star & -\bar{\mathrm{y}}^2
            \end{bmatrix} 
            \!+\!
            \begin{bmatrix}
                \mathrm{S} & 0 \\ 0 & -1
            \end{bmatrix}
            \succeq 0, \label{eq:terminal constraints OP 5}
            \\
            & {\color{white} \mathrm{s.t.}} \qquad
            \mxS \preceq \mxS_{t-1}, \qquad \mathrm{if} \  t \geq 1. \label{eq:terminal constraints OP 6}
        \end{align}
    \end{subequations}
and $\mathrm{S}_t$ is selected as the optimal solution of this optimization program. The last inequality \eqref{eq:terminal constraints OP 6} ensures that $\mathcal{R}_{t-1} \subseteq \mathcal{R}_t$. At $t=0$, \eqref{eq:terminal constraints OP 6} is omitted, and the initial terminal set $\mathcal{R}_0$ is determined by solving the remaining optimization problem. As the robust parameter set contracts over time, the constraints in \eqref{eq:terminal constraints} and their equivalent LMI formulation \eqref{eq:terminal constraints LMI} become less restrictive. Consequently, the previous solution $\mxS_{t-1}$ remains feasible for the current optimization problem, allowing the additional constraint $\mxS_t \preceq \mxS_{t-1}$ to be imposed without compromising feasibility.

\begin{remark}
    \normalfont
    The preceding development of robust parameter sets, safety constraints, and online terminal sets is independent of the choice of objective function. Therefore, the proposed framework can be applied to different control tasks by selecting an appropriate objective function, without modifying the safety and recursive feasibility guarantees.
\end{remark}

\subsection{Control Objective Function}\label{sec:Objective function}
The proposed scheme can be applied to a broad class of control tasks by appropriately selecting the exploitation cost $J$ introduced in Section~\ref{sec:Online Input Design Scheme} to characterize the desired control objective.
Nevertheless, due to the importance of tracking in practice, we specifically consider two relevant objective functions in the sequel, that share the same exploration term but differ in their exploitation objectives, each designed to find the optimal input $\vcu_t$ by minimizing the exploitation cost over the uncertain parameter set. 
The first objective function uses a worst-case cost, which seeks the control input that minimizes the maximum possible exploitation cost under all parameters in the robust parameter set. The second objective function uses an expected cost, which instead minimizes the average exploitation cost with respect to the probability distribution of the uncertain parameters. For ease of discussion, in the rest of this paper, we use a tracking task as our control task, i.e., given a reference $(z_t)_{t \in \mathbb{Z}_+}$, the exploitation cost at time $t$ is $\ell_t = \|\mu^\tr \varphi_t - z_t\|^2$. However, the proposed framework is not restricted to tracking cost functions and can be extended to general quadratic functions.

\subsubsection{Tracking Control: Worst-case Objective Function}
Given reference sequences $(z_t)_{t \in \mathbb{Z}_+}$, the worst-case objective function is formulated as
\begin{equation}\label{eq:cost function worst-case}
    L_{\mathrm{wc}} = \min_{\{\mathrm{u}_{k|t}\}_{k=0}^\nsP} \max_{\nu \in \mathcal{C}_t} \sum_{k=0}^\nsP \left( \|z_{k+t} - \nu^\tr \varphi_{k|t}\|^2 \right),
\end{equation}
where the inner maximization captures the worst-case tracking error over all parameters $\nu$ in $\mathcal{C}_t$. 
The outer minimization then finds the control input that minimizes this worst-case tracking error. The resulting min-max problem cannot be directly handled by standard optimization tools. Consequently, we derive the following equivalent convex formulation.

\begin{theorem}
    The worst-case tracking error \eqref{eq:cost function worst-case} can be equivalently reformulated as the optimal value of the following SDP problem:
    \begin{subequations}\label{eq:cost function convex LMI}
\begin{align}
    &\min_{\lambda \geq 0, c, \{\mathrm{u}_{k|t}, \mathrm{D}_{k|t}\}_{k=0}^\nsP} 
    \    c \nonumber
    \\
    & 
    \quad \  \qquad \mathrm{s.t.}
    \qquad \  \quad
    \begin{bmatrix}
        \mathrm{D}_{\nsP|t} & -\!\lambda d_t \!+\! \sum_{k=0}^\nsP z_{k+t}\varphi_{k|t}\\
        \star & \lambda a_t \!-\! \lambda \beta_t\!+\! c \!-\! \bar{z}_t
    \end{bmatrix}
    \succeq 0, \label{eq:cost function convex LMI 4}
    \\
    &
    \qquad \qquad  \qquad \quad  \quad
    \begin{bmatrix}
        \lambda \mathrm{A}_t  - \mathrm{D}_{0|t} & \varphi_{0|t}\\
        \star & 1
    \end{bmatrix}    
    \succeq 0, \label{eq:cost function convex LMI 3}
    \\
    & 
    \qquad \qquad  \qquad \quad \quad
    \begin{bmatrix}
        \mathrm{D}_{k|t} \!-\! \mathrm{D}_{k+1|t} & \varphi_{k+1|t}\\
        \star & 1
    \end{bmatrix}
    \succeq 0,  \label{eq:cost function convex LMI 2} 
    \\  
    & \qquad \qquad \qquad \qquad  \qquad \qquad \qquad \quad \  
    \forall k \in  \mathbb{Z}_{\nsP-1}, \nonumber
\end{align}
\end{subequations}
where $\bar{z}_t := \sum_{k=0}^\nsP z_{k+t}^2$ while $d_t$ and $a_t$ are defined under \eqref{eq:biggest nesting set}.
\end{theorem}
\begin{proof}
    We first find the dual problem for the inner optimization.
    Define
\begin{equation}\label{eq:cost function max}
\begin{split}
    c :&= \max_{\nu \in \mathcal{C}_t} \sum_{k=0}^\nsP \|\nu^\tr \varphi_{k|t} - z_{k+t}\|^2\\
    &= \max_{\nu \in \mathcal{C}_t} \sum_{k=0}^\nsP \nu^\tr \varphi_{k|t} \varphi_{k|t}^\tr \nu - 2 z_{k+t} \varphi_{k|t}^\tr \nu + z_{k+t}^2.
\end{split}
\end{equation}
The above equation represents the maximal cost of the squared tracking error. The maximization in \eqref{eq:cost function max} has a quadratic form in $\nu$ constrained over the robust parameter set $\mathcal{C}_t$. Hence, applying the S-procedure, the original min-max problem \eqref{eq:cost function worst-case} can be equivalently transformed as 
\begin{equation}\label{eq:cost function min_max S-procedure}
\begin{split}
    L_{\mathrm{wc}} =  &\min_{\lambda \geq 0, c, \{\mathrm{u}_{k|t}\}_{k=0}^\nsP} 
     \qquad c 
     \\
    &\  \qquad \mathrm{s.t.} \quad \quad \  \qquad 
    \lambda\begin{bmatrix}
        \mathrm{A}_t & -d_t\\
        \star & a_t - \beta_t
    \end{bmatrix}
     -
    \sum_{k=0}^\nsP \begin{bmatrix}
        \varphi_{k|t} \varphi_{k|t}^\tr & -z_{k+t}\varphi_{k|t}\\
        \star & z_{k+t}^2 - \frac{c}{\nsP+1}
    \end{bmatrix}
    \!\succeq\! 0,\\
\end{split}\!\!\!\!\!\!\!\!\!
\end{equation}
where $\lambda$ is a non-negative Lagrange multiplier.
Similarly to \eqref{eq:safety constraints convex LMI}, the above optimization program can be reformulated to a convex program. However, to address the nonlinearity of the summarization of $\varphi_{k|t}\varphi_{k|t}^\tr$, a set of auxiliary matrix variables $\{\mathrm{D}_{k|t}\}_{k=0}^\nsP$ is designed as 
\begin{equation}\label{eq:D_k|t}
    \begin{split}
        \lambda \mathrm{A}_t - \varphi_{0|t}\varphi_{0|t}^\tr &\succeq \mathrm{D}_{0|t},\\
        \mathrm{D}_{0|t} - \varphi_{1|t}\varphi_{1|t}^\tr &\succeq \mathrm{D}_{1|t},\\
        \vdots\\
        \mathrm{D}_{\nsP-1|t} - \varphi_{\nsP|t}\varphi_{\nsP|t}^\tr &\succeq \mathrm{D}_{\nsP|t}.\\
    \end{split}
\end{equation}
Each equation above can be reformulated to a LMI using Schur complement as
\begin{equation}
    \begin{bmatrix}
        \lambda \mathrm{A}_t  - \mathrm{D}_{0|t} & \varphi_{0|t}\\
        \star & 1
    \end{bmatrix}    \succeq 0,
\end{equation}
or
\begin{equation}
    \begin{bmatrix}
        \mathrm{D}_{k|t} - \mathrm{D}_{k+1|t} & \varphi_{k+1|t}\\
        \star & 1
    \end{bmatrix}    \succeq 0,
\end{equation}
for $k \in \mathbb{Z}_{\nsP-1}$.
Consequently, the optimization program \eqref{eq:cost function min_max S-procedure} is equivalent to \eqref{eq:cost function convex LMI}. This completes the proof.
\end{proof}

Thus, the original min-max objective function in \eqref{eq:cost function worst-case} is equivalently transformed into a convex optimization problem augmented with LMIs.

\begin{remark}
\normalfont
    The above objective function \eqref{eq:cost function convex LMI} focus on exploitation, i.e., minimizing the tracking error based on the current model estimate. However, in many applications it is also desirable to promote exploration, encouraging the controller to actively gather informative data and improve parameter estimation over time. This active learning component can be incorporated by introducing an exploration term of the form 
    \begin{equation}\label{eq:active learning regularization}
        \sum_{k=0}^{\nsR} - \mathrm{tr} (\tilde{\mathrm{A}}_t + \frac{1}{\sigma_w^2}\varphi_{k|t} \varphi_{k|t}^\tr),
    \end{equation}
    where $\nsR \leq \nsP$ represents the exploration horizon and $\tilde{\mathrm{A}}_t$ arises from the update of the robust parameter set, as described in \eqref{eq:bayesian linear regressor}. This exploration term promotes an increase in the trace of the predicted information matrix, thereby encouraging inputs that excites to actively learn the unknown parameters.
\end{remark}
Since $\tilde{\mathrm{A}}_t$ is given at time $t$, we can omit it in the cost function. The min-max objective function with active learning regularization term can thus be formulated as
\begin{subequations}\label{eq:cost function convex LMI active}
\begin{align}
    L_{\mathrm{wc}}^{\mathrm{(ac)}} = 
    &\min_{\lambda \geq 0, c, \{\mathrm{u}_{k|t}, \mathrm{D}_{k|t}\}_{k=0}^\nsP} 
    \ 
    c 
    - 
    \sum_{k=0}^{\nsR}  \omega \|\varphi_{k|t}\|^2 &\label{eq:cost function convex LMI active 1}
    \\
    & 
    \quad \  \qquad \mathrm{s.t.}
    \qquad \ \quad
    \begin{bmatrix}
        \mathrm{D}_{\nsP|t} & -\!\lambda d_t \!+\! \sum_{k=0}^\nsP z_{k+t}\varphi_{k|t}\\
        \star & \lambda a_t \!-\! \lambda \beta_t\!+\! c \!-\! \bar{z}_t
    \end{bmatrix}
    \succeq 0, \label{eq:cost function convex LMI active 4}\\
    &
    \qquad \qquad  \qquad \quad \quad
    \begin{bmatrix}
        \lambda \mathrm{A}_t  - \mathrm{D}_{0|t} & \varphi_{0|t}\\
        \star & 1
    \end{bmatrix}    
    \succeq 0, \label{eq:cost function convex LMI active 3}\\
    &
    \qquad \qquad  \qquad \quad \quad
    \begin{bmatrix}
        \mathrm{D}_{k|t} - \mathrm{D}_{k+1|t} & \varphi_{k+1|t}\\
        \star & 1
    \end{bmatrix}
    \succeq 0,  \label{eq:cost function convex LMI active 2}
    \\  
    & \qquad \qquad \qquad \qquad  \ \, 
    \forall k \in  \mathbb{Z}_{\nsP-1}, \nonumber
\end{align}
\end{subequations}
where $\omega$ is a tuning parameter.
Notice that the resulting optimization program \eqref{eq:cost function convex LMI active} is not convex due to the exploration term in the cost function. In Section \ref{sec:Convex Approximation}, we introduce an approximation method to formulate a convex program that yields a lower-bound solution to optimization program \eqref{eq:cost function convex LMI active}.
\begin{remark}
\normalfont
     In practice, $\omega$ can be set as inversely proportional to $\lambda_\mathrm{\max}(\tilde{\mathrm{A}}_t)$, i.e., $\omega = c_0/\lambda_\mathrm{\max}(\tilde{\mathrm{A}}_t)$, where $c_0$ is a positive constant. Thus, the optimization program motivates exploration when the uncertainty set is large while focusing more on exploitation when the uncertainty set is small.
\end{remark}

\subsubsection{Tracking Control: Expectation Objective Function}\label{sec:Expectation Objective Function}
Another candidate objective function is based on the expected tracking error with respect to the uncertain parameter as
\begin{equation}\label{eq:cost function expected}
    L_{\mathrm{exp}} = \min_{\{\mathrm{u}_{k|t}\}_{k=0}^\nsP} \mathds{E}\left[ \sum_{k=0}^\nsP \left( \|z_{k+t} - \nu^\tr \varphi_{k|t}\|^2 \right) \right].
\end{equation}
From \eqref{eq:bayesian linear regressor}, the posterior distribution of the unknown parameter $\mu$ is $\mathcal{N}(\tilde{{\mu}}_t, \tilde{\Sigma}_t)$. Using this distribution, the objective function \eqref{eq:cost function expected} can be expanded as
\begin{equation}
    \begin{split}
        L_{\mathrm{exp}} &= \min_{\{\mathrm{u}_{k|t}\}_{k=0}^\nsP} \mathds{E}\Bigg[ \sum_{k=0}^\nsP \Big( z_{k+t}^2 - 2z_{k+t} \nu^\tr \varphi_{k|t} 
         + \varphi_{k|t}^\tr \nu \nu^\tr \varphi_{k|t} \Big) \Bigg]\\
        &= \min_{\{\mathrm{u}_{k|t}\}_{k=0}^\nsP}  \sum_{k=0}^\nsP \Bigg( z_{k+t}^2 - 2z_{k+t} \tilde{\mu}_t^\tr \varphi_{k|t} 
        + \varphi_{k|t}^\tr \left( \tilde{\Sigma}_t + \tilde{\mu}_t \tilde{\mu}_t^\tr \right) \varphi_{k|t} \Bigg).
    \end{split}
\end{equation}
This objective function is also convex and can be efficiently solved online. In contrast to the worst-case criterion in \eqref{eq:cost function worst-case}, which optimizes performance under the most adverse parameter realization, the expected-cost formulation focuses on average-case performance. By incorporating the posterior distribution of the uncertain parameter, this objective reflects the typical tracking behavior rather than the worst possible scenario. Moreover, the exploration term in \eqref{eq:active learning regularization} can also be included to lead an active learning and balance exploitation and exploration, as 

\begin{equation}
    \begin{split}
        L_{\mathrm{exp}}^{\mathrm{(ac)}}
        &= \min_{\{\mathrm{u}_{k|t}\}_{k=0}^\nsP} \sum_{k=0}^\nsP \Bigg( z_{k+t}^2  + \varphi_{k|t}^\tr \left( \tilde{\Sigma}_t + \tilde{\mu}_t \tilde{\mu}_t^\tr \right) \varphi_{k|t}
        - 2z_{k+t} \tilde{\mu}_t^\tr \varphi_{k|t}\Bigg) 
        - \sum_{k=0}^{\nsR} \omega \|\varphi_{k|t}\|^2 . 
    \end{split}
\end{equation}
Note that the above objective function remains convex when the weighting parameter $\omega$ is small. In particular, convexity is guaranteed if
\begin{equation}
    \omega < \frac{\min(\nsP, n_{\mathrm{g}} n_{\mathrm{u}})}{\nsR} \lambda_{\min} (\Sigma_t + \mu_t \mu_t^\tr).
\end{equation}
Under this condition, the associated Hessian matrix remains positive semidefinite, thereby ensuring convexity of the objective function.

\subsection{Online Performance-aware Safety-constrained Input Design Algorithm}
Now we summarized to the optimization problem and formulate the algorithm. Define the control variable  $v_{\mathrm{wc}} = \{\lambda, c, \{\!\mathrm{u}_{k|t}, \bar{\lambda}_{k}, \mathrm{D}_{k|t},  \bar{\mathrm{D}}_{k|t}\!\}_{k=0}^\nsP\}$. Using the worst-case objective function, the optimization program can be formulated as
\begin{subequations}\label{eq:MPC worst-case}
\begin{align}
    L_{\mathrm{wc}}^{\mathrm{(ac)}} \ = \  
    &\min_{v_{\mathrm{wc}}} 
    \qquad\!\! c 
    -  
    \sum_{k=0}^{\nsR}  \omega \|\varphi_{k|t}\|^2   \label{eq:MPC worst-case 1}\\
    &
    \ \ \,\!\mathrm{s.t.}\!\!\!\!
    \qquad
    \begin{bmatrix}
        \mathrm{D}_{\nsP|t} & -\!\lambda d_t \!+\! \sum_{k=0}^\nsP z_{k+t}\varphi_{k|t}\\
        \star & \lambda a_t-\! \lambda \beta_t\!+\! c \!-\! \bar{z}_t
    \end{bmatrix}
    \!\succeq \!0,  \label{eq:MPC worst-case 2} \\
    &
    \qquad \quad \ 
    \begin{bmatrix}
        \lambda \mathrm{A}_t  - \mathrm{D}_{0|t} & \varphi_{0|t}\\
        \star & 1
    \end{bmatrix}    
    \!\succeq \!0, \label{eq:MPC worst-case 3}\\
    &
    \qquad \quad \ 
    \begin{bmatrix}
        \mathrm{D}_{k_1|t} - \mathrm{D}_{k_1+1|t} & \varphi_{k_1+1|t}\\
        \star & 1
    \end{bmatrix}
    \!\succeq \!0, \label{eq:MPC worst-case 4}\\
    &
    \qquad \quad \ 
    \begin{bmatrix}
            \bar{\mathrm{D}}_{k_2|t} &  -\bar{\lambda}_{k_2|t} d_t\\
            \star & \bar{\lambda}_{k_2|t} a_t -\! \bar{\lambda}_{k_2|t} \beta_t \!+\! \bar{\mathrm{y}}^2 
    \end{bmatrix}
    \!\succeq \!0, \label{eq:MPC worst-case 5}\\
    &
    \qquad \quad \ 
    \begin{bmatrix}
            \bar{\lambda}_{k_2|t}\mathrm{A}_t - \bar{\mathrm{D}}_{k_2|t} & \varphi_{k_2|t}\\
            \star & 1
        \end{bmatrix}
    \!\succeq \!0, \label{eq:MPC worst-case 6}\\
    &
    \qquad \quad \ 
    \varphi_{\nsP|t}^\tr \mathrm{S}_t \varphi_{\nsP|t}  \!\leq\! 1, \label{eq:MPC worst-case 7}
    \\
    &
    \qquad \quad \ 
    \|\mathrm{u}_{k|t}\| \leq \bar{\mathrm{u}},
    \\
    &
    \qquad \quad \ 
    \bar{\lambda}_{k_2|t} \!\geq\! 0,
    \\
    &
    \qquad \quad \ 
    \lambda \!\geq\! 0,
    \\
    &
    \qquad \quad \ 
    \forall k_1 \in \mathbb{Z}_{\nsP\!-\!1}, 
    \ 
    \forall k_2 \in \mathbb{Z}_{\nsP}. \nonumber
\end{align}
\end{subequations}
Define $v_{\mathrm{exp}} = \{\mathrm{u}_{k|t}, \bar{\lambda}_{k|t}, \bar{\mathrm{D}}_{k|t}\}_{k=0}^{\nsP}$. Similarly, using the expectation cost described in Section \ref{sec:Expectation Objective Function}, the optimization program can be formulated as
\begin{subequations}\label{eq:MPC Expectation}
\begin{align}
     L_{\mathrm{exp}}^{\mathrm{(ac)}} \  = \ 
     &\min_{v_{\mathrm{exp}}} 
    \quad \sum_{k=0}^\nsP \Big( z_{k+t}^2 - 2z_{k+t} \tilde{\mu}_t^\tr \varphi_{k|t} + 
    \varphi_{k|t}^\tr ( \tilde{\Sigma}_t + \tilde{\mu}_t \tilde{\mu}_t^\tr ) \varphi_{k|t} \Big)
        \!-\! \sum_{k=0}^{\nsR}  \omega \|\varphi_{k|t}\|^2\label{eq:MPC Expectation 1}\\
    & \ \  
    \mathrm{s.t.}
    \quad \ 
    \begin{bmatrix}
            \bar{\mathrm{D}}_{k|t} &  -\bar{\lambda}_{k|t} d_t\\
            \star & \bar{\lambda}_{k|t} a_t \!-\! \bar{\lambda}_{k|t} \beta_t \!+\! \bar{\mathrm{y}}^2 
    \end{bmatrix}
    \succeq 0, \label{eq:MPC Expectation 2}\\
    &
    \qquad \quad \ 
    \begin{bmatrix}
            \bar{\lambda}_{k|t}\mathrm{A}_t - \bar{\mathrm{D}}_{k|t} & \varphi_{k|t}\\
            \star & 1
        \end{bmatrix}
        \succeq 0, \label{eq:MPC Expectation 3} \\
    &
    \qquad \quad \ 
    \varphi_{\nsP|t}^\tr \mathrm{S}_t \varphi_{\nsP|t} \leq 1, \label{eq:MPC Expectation 4} \\
    &
    \qquad \quad \ 
    \|\mathrm{u}_{k|t}\| \leq \bar{\mathrm{u}},
    \\
    &
    \qquad \quad \ 
    \bar{\lambda}_{k|t} \!\geq\! 0, 
    \quad \forall k \in \mathbb{Z}_\nsP.
\end{align}
\end{subequations}
At each time step, after solving the above optimization program, the parameter set is updated using Algorithm \ref{Al:Update Robust Parameter Set}. The resulting safe online input design algorithm is summarized in Algorithm \ref{Al: SID}. 

\begin{algorithm}[t]
    \caption{Safe Performance-aware Online Input Design}\label{Al: SID}
    \begin{algorithmic}
        \Statex \textbf{Input:} $\mathrm{A}_{0}, \mu_{0}, \beta_{0}, \{\mathrm{z}\}_{k=0}^T$.
        \Statex \textbf{Output:} $\mathrm{A}_{T}, \mu_{T}, \beta_{T}$.
            \State $i \gets 0, t \gets 0$.
            \State $\tilde{\mathrm{A}}_0 \gets \mathrm{A}_0$, $\tilde{\Sigma}_0 \gets \mathrm{A}_0^{-1}$ and $\tilde{\mu}_0 \gets \mu_0$.
            \While {$t<=T$}
                \State Solve \eqref{eq:MPC worst-case} or \eqref{eq:MPC Expectation}.
                \State $\vcu_t \gets 
                    \vcu_{0|t}^*$.
                \State  Apply $\vcu_t$ and obtain $\vcy_t$.
                \State Update parameter set for time $t+1$ using Algorithm \ref{Al:Update Robust Parameter Set}.
                \State $t \gets t+1$.
            \EndWhile
    \end{algorithmic}
\end{algorithm}

\begin{theorem}[Recursive Feasibility]\label{thm:recursive feasibility}
    Assume that the optimization program \eqref{eq:MPC worst-case} (or \eqref{eq:MPC Expectation}) is feasible at time 0, and the initial online terminal set is not empty, i.e., \eqref{eq:terminal constraints OP} is feasible at time 0. Let the robust parameter set be updated according to \eqref{eq:robust parameter set update}. Let the online terminal set also be updated according to \eqref{eq:terminal constraints OP}. Then, the optimization program \eqref{eq:MPC worst-case} (or \eqref{eq:MPC Expectation}) is feasible for all $t \in \mathbb{Z}_+$.
\end{theorem}
\begin{proof}
    The proof is in Appendix~\ref{app:proof of recursive feasibility}.
\end{proof}


\section{Tractability Promotion via Convex Approximation}\label{sec:Convex Approximation}
As discussed in Section \ref{sec:Objective function}, the cost function \eqref{eq:MPC worst-case 1} is nonconvex due to the presence of the exploration term. In practice, one can solve the optimization problem through a nonlinear solver. To reduce the online computational burden, we introduce a convex approximation of the optimization problem.
\subsection{Convex Formulation}\label{sec:Convex Formulation}
Consider the worst-case objective function with exploration term 
\begin{equation}\label{eq:cost function worst-case active learning}
\begin{split}
    L_{\mathrm{wc}}^{\mathrm{(ac)}} = \min_{\mathrm{u}_{0|t},\cdots, \mathrm{u}_{\nsP|t}} \max_{\nu \in {\mathcal{C}}_t} \sum_{k=0}^\nsP \Big( \|&z_{k+t} - \nu^{\mathsf T} \varphi_{k|t}\|^2\Big) 
    - \sum_{k=0}^{\nsR} \omega \|\varphi_{k|t}\|^2.
\end{split}
\end{equation}
Define 
\begin{equation}\label{eq:s_t}
    s_t := \sum_{k=0}^{\nsR} \varphi_{k|t}^{\mathsf T}\varphi_{k|t},
\end{equation}
and the maximal value of the inner optimization problem as
\begin{equation}\label{eq:approx active cost}
    \tilde{c} := \max_{\nu \in {\mathcal{C}}_t} \sum_{k=0}^\nsP  \left( \nu^{\mathsf T} \varphi_{k|t} \varphi_{k|t}^{\mathsf T} \nu - 2 z_{k+t} \varphi_{k|t}^{\mathsf T} \nu + z_{k+t}^2 \right) - \omega s_t,
\end{equation}
The above equation is similar to \eqref{eq:cost function max}, but we add the exploration term in it. Therefore, analogous to \eqref{eq:cost function convex LMI}, using the S-procedure again, the min-max problem \eqref{eq:cost function worst-case active learning} is equivalent to 
\begin{subequations}\label{eq:cost function convex LMI active learning}
\begin{align}
    L_{\mathrm{wc}}^{\mathrm{(ac)}} = 
    & \min_{\lambda, \tilde{c}, \{\mathrm{u}_{k|t}, \mathrm{D}_{k|t}\}_{k=0}^\nsP} 
    \quad  \tilde{c}  \qquad \qquad &\label{eq:cost function convex LMI active learning 1}\\
    & \qquad \quad
    \mathrm{s.t.}
    \qquad \quad 
    \begin{bmatrix}
        \mathrm{D}_{\nsP|t} & - \lambda d_t + \sum_{k=0}^\nsP z_{k+t}\varphi_{k|t}\\
        * & \lambda a_t - \!\lambda \beta_t \!+\! \tilde{c} \!- \bar{z}_t +\! \omega s_t
    \end{bmatrix}
    \!\succeq\! 0,  \label{eq:cost function convex LMI active learning 2} \\
    & \qquad \qquad \qquad \quad  \ \   
    \eqref{eq:cost function convex LMI active 3}, \eqref{eq:cost function convex LMI active 4}, \eqref{eq:s_t}, \ 
    \forall k \in \mathbb{Z}_{\nsP-1}.\nonumber
\end{align}
\end{subequations}
This reformulation is mathematically equivalent to the original formulation \eqref{eq:cost function convex LMI active}, with the nonconvex exploration term transferred from the objective function to the constraints. In the above problem, the only nonconvex component is the equality constraint of $s_t$ in \eqref{eq:s_t}, which is equivalent to,
\begin{subequations}\label{eq:s_t inequality constraints}
\begin{align}
        s_t & \geq \sum_{k=0}^{\nsR} \varphi_{k|t}^{\mathsf T} \varphi_{k|t}, \label{eq:s_t inequality constraints 1}\\
        s_t & \leq \sum_{k=0}^{\nsR} \varphi_{k|t}^{\mathsf T} \varphi_{k|t}. \label{eq:s_t inequality constraints 2}
\end{align}
\end{subequations}
Constraint \eqref{eq:s_t inequality constraints 1} is convex. Constraint \eqref{eq:s_t inequality constraints 2} can be relaxed by summing \eqref{eq:D_k|t}, yielding
\begin{equation}
    \lambda \mathrm{A}_t - \mathrm{D}_{\nsR|t} \succeq \sum_{k=0}^{\nsR} \varphi_{k|t}\varphi_{k|t}^{\mathsf T}, 
\end{equation}
which implies that 
\begin{equation}
    \mathrm{tr} \left(\lambda \mathrm{A}_t - \mathrm{D}_{\nsR|t} \right) \geq \sum_{k=0}^{\nsR} \varphi_{k|t}^{\mathsf T} \varphi_{k|t}.
\end{equation}
Hence, we can use the above equation to relax the constraint \eqref{eq:s_t inequality constraints 2}. The optimization problem \eqref{eq:cost function convex LMI active learning} can be approximated as 
\begin{subequations}\label{eq:cost function convex LMI active learning app}
\begin{align}
    \underline{L}_{\mathrm{wc}}^{\mathrm{(ac)}} = 
    &\min_{\lambda, \tilde{c}, \{\mathrm{u}_{k|t}, \mathrm{D}_{k|t}\}_{k=0}^\nsP} 
    \tilde{c} \qquad \quad & \label{eq:cost function convex LMI active learning app 1}\\
    &\qquad \quad 
    \mathrm{s.t.}
    \qquad \  
    \begin{bmatrix}
        \mathrm{D}_{\nsP|t} & - \lambda d_t + \sum_{k=0}^\nsP z_{k+t}\varphi_{k|t}\\
        * & \lambda a_t \!- \!\lambda \beta_t \!+\! \tilde{c} \!- \bar{z}_t +\! \omega s_t
    \end{bmatrix}
    \succeq  0,  \label{eq:cost function convex LMI active learning app 2} \\
    & \qquad \qquad \qquad \quad   
    \sum_{k=0}^{\nsR} \varphi_{k|t}^{\mathsf T} \varphi_{k|t} 
    \leq s_t, \label{eq:cost function convex LMI active learning app 3}  \\
    &\qquad \qquad \qquad \quad 
    \mathrm{tr} \left(\lambda \mathrm{A}_t - \mathrm{D}_{\nsR|t} \right) 
    \geq  s_t  \label{eq:cost function convex LMI active learning app 4} \\ 
    &\qquad \qquad \qquad \quad 
    \eqref{eq:cost function convex LMI active 3}, \eqref{eq:cost function convex LMI active 4}, 
    \ 
    \forall k \in \mathbb{Z}_{\nsP-1}.\nonumber
\end{align}
\end{subequations}

\begin{remark}
\normalfont
    The above optimization problem relaxes the upper bound constraint of $s_t$. It follows from the LMI constraint \eqref{eq:cost function convex LMI active learning app 2} that \eqref{eq:cost function convex LMI active learning app} provides a lower bound on the optimal value of the original problem \eqref{eq:cost function convex LMI active learning}, i.e., $\underline{L}_{\mathrm{wc}}^{\mathrm{(ac)}} \leq L_{\mathrm{wc}}^{\mathrm{(ac)}}$.
\end{remark}
\begin{remark}
\normalfont
It should be noted that the above convexification procedure is applicable to any control task, provided that a convex lower bound for the exploitation term is available.
\end{remark}

\subsection{Experiment Analysis}\label{sec:Experiment Analysis}
\begin{figure}[t!]
   \centering
   \includegraphics[trim={0cm 0cm 0cm 0cm},clip]{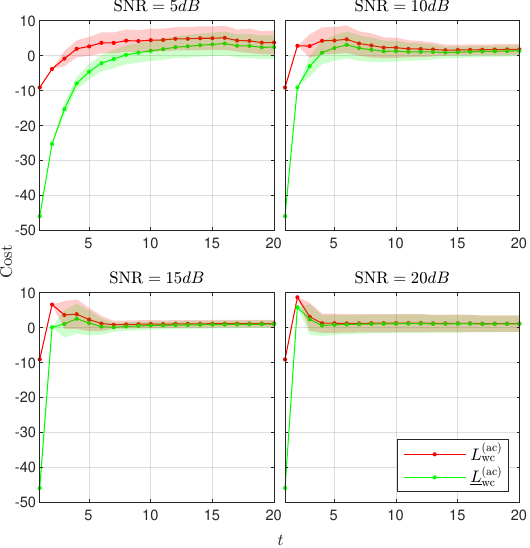}
   \caption{Evolution of the cost function under four SNR levels. Results are averaged over 100 Monte Carlo realizations. The dotted lines represent the sample mean, and the shaded regions represent the mean ± one standard deviation. }
   \label{img:convex_approx}
\end{figure}
A numerical experiment is conducted to evaluate the convex approximation of the optimization problem. In this experiment, the truncation parameter of the system is set to $n_\mathrm{g} = 2$. The initial mean and standard deviation of both parameters are specified as 1 and 0.6, respectively. The real parameter is $\mu = \begin{bmatrix} 1.5 & 0.2 \end{bmatrix}^{\mathsf T}$. A constant tracking reference is applied, given by $z_t = 1.5$ for all $t$. The input and output constraints are defined as $|u_t| < 1$ and $|y_t| < 4$, respectively. The noise variance is selected such that the corresponding SNR is approximately $5$, $10$, $15$, and $20\mathrm{dB}$. The performance of the proposed Algorithm 2, based on the worst-case cost function, is evaluated by comparing the formulations in \eqref{eq:cost function convex LMI active} and \eqref{eq:cost function convex LMI active learning app}. The prediction horizons are chosen as $\nsP = 4$ and $\nsR = 2$, respectively. The total simulation length is set to 50 steps. For each SNR level, $N_{\mathrm{MC}} = 100$ Monte Carlo simulations are conducted using independently generated Gaussian noise sequences.

Figure \ref{img:convex_approx} shows the results. For all considered noise levels, the cost is initially low, increases rapidly during the early stage, and then gradually approaches a steady level. This behavior can be attributed to the exploration-exploitation trade-off inherent in both methods. In the early stage, the algorithms prioritize exploration to improve parameter estimation, which introduces additional excitation into the system and consequently increases the cost. As the estimation accuracy improves over time, both methods progressively shift toward exploitation, emphasizing tracking performance rather than information acquisition. This transition reduces the contribution of the exploration term to the objective. Nevertheless, mild exploration persists throughout the process, which contributes to a gradual reduction in the tracking error over time until a steady behavior is achieved. Moreover, the exploration phase converges more rapidly at higher SNR levels, as improved measurement quality accelerates parameter learning.

A comparison of the two methods shows that the cost associated with $\underline{L}_{\mathrm{wc}}^{\mathrm{(ac)}}$ is consistently lower than that of $L_{\mathrm{wc}}^{\mathrm{(ac)}}$, as expected. In the initial stage, exploration dominates the control strategy, leading to a relatively large gap between the two cost values. As time progresses, the tracking error becomes the dominant component of the cost, reducing the influence of exploration and resulting in a more accurate approximation. The average computation time for the convex approximation \eqref{eq:cost function convex LMI active learning app} is approximately $0.26s$, compared with $26.64s$ for the original problem \eqref{eq:cost function convex LMI active}.

Overall, the experimental results demonstrate that the proposed approximation significantly reduces the online computational burden compared with the original formulation. Additionally, the approximation accuracy improves over time, with the lower-bound approximation becoming tighter as the two cost values approach each other.

\section{Safe Input Design for MIMO System}\label{sec:Extension to MIMO System}
In this section, we extend the proposed safe input design framework to MIMO systems. We first introduce the prior knowledge and robust parameter sets for the MIMO setting, and then formulate the corresponding worst-case and expectation-based objective functions.
\subsection{Prior Knowledge and Robust Parameter Set}\label{sec:Prior Knowledge and Robust Parameter Set}
Assume now that the input and output of the system are $\mathrm{u} \in \mathbb{R}^{n_\mathrm{u}}$ and  $\mathrm{y} \in \mathbb{R}^{n_\mathrm{y}}$, respectively. Consequently, the impulse response of the system is $\{\mathrm{G}_t\}_{t=0}^{n_\mathrm{G}}$, where $ \mathrm{G}_t \in \mathbb{R}^{n_{\mathrm{y}} \times n_{\mathrm{u}}}$. 
Using the notations in \eqref{eq:definition mu} and \eqref{eq:definition phi}, and recalling that $\mu^i$ denotes the transpose of the i-th row of $\mu$, we have
\begin{equation}
    \mathrm{y}^{(i)} = \mu^{i^{\mathsf T}} \varphi_t + \mathrm{w}_t^{(i)},
\end{equation}
where $\mathrm{y}^{(i)}$ and $\mathrm{w}^{(i)}$ denote the i-th row of $\mathrm{y}$ and $\mathrm{w}$, respectively.
We assume that for each row $i$, the prior distribution of $\mu^i$ is a Gaussian, i.e.,
\begin{equation}
    \mu^{i} \sim \mathcal{N}({\mu}^{i}_0, {\Sigma}^{i}_0),
\end{equation}
and $\mu^i$ are mutually independent.

For each row i, Algorithm \ref{Al:Update Robust Parameter Set} can be applied to update the robust parameter set. Consequently, we can obtain the posterior distribution of the unknown parameter $\mu^i$ as $\mathcal{N}(\tilde{{\mu}}_t^i, \tilde{\Sigma}_t^i)$ and a sequence of ellipsoids ${\mathcal{C}}_{t}^{i} := \{\nu : \|\nu - {\mu}_t^{i}\|_{{\Sigma}_t^{i^{-1}}} \leq ({\beta}_t^i)^\frac{1}{2} \}$ as a robust parameter set of $\mu^i$ at the time instant $t$. Using the robust parameter set ${\mathcal{C}}_{t}^{i}$, one can reformulate the safety constraints using \eqref{eq:safety constraints convex LMI} for each row.

\begin{remark}
\normalfont
    The output safety set is assumed to be a hyperrectangle for computational tractability. More general safety sets with nonempty interior can be conservatively approximated by an inscribed hyperrectangle, allowing the proposed framework to be applied while preserving safety guarantees. Although this approximation may introduce additional conservatism, its effect on control performance can diminish as parameter uncertainty decreases through online learning.
\end{remark}

\subsection{Control Objective Function}\label{sec:Objective Function MIMO}
\subsubsection{Tracking Control: Worst-case Objective Function}\label{sec:Worst-case Objective Function MIMMO}
The worst-case objective function for the MIMO system is defined as
\begin{equation}\label{eq:cost function worst-case MIMO}
\begin{split}
    L_{\mathrm{wc}} = \min_{\{\mathrm{u}_{k|t}\}_{k=0}^\nsP} \max_{\substack{\nu^i \in {\mathcal{C}}^i_t, \\ i \in \mathbb{N}_{n_\vcy}}} 
    &\sum_{k=0}^\nsP   \sum_{i=1}^{n_{\mathrm{y}}} \left(\|z_{k+t}^{(i)} - \nu^{i^{\mathsf T}} \varphi_{k|t}\|^2 \right) 
    - \sum_{k=0}^{\nsR} \omega n_{\mathrm{y}} \|\varphi_{k|t}\|^2.
\end{split}
\end{equation}
Compared with \eqref{eq:cost function worst-case active learning}, the objective function additionally sums over the output dimensions. The constant $n_{\mathrm{y}}$ in the last term is added for simplicity and can be absorbed into $\omega$. 
Thus, we can reformulate the above equation to
\begin{equation}\label{eq:cost function worst-case MIMO 2}
\begin{split}
    L_{\mathrm{wc}} \!=\! \min_{\{\mathrm{u}_{k|t}\}_{k=0}^\nsP} \! \max_{\substack{\nu^i \in {\mathcal{C}}^i_t, \\ i \in \mathbb{N}_{n_\vcy}}}
    \sum_{i=1}^{n_{\mathrm{y}}} \Bigg( \sum_{k=0}^\nsP  
    \Big(\| & z_{k+t}^{(i)} - \nu^{i^{\mathsf T}} \varphi_{k|t}\|^2\Big) 
    - \sum_{k=0}^{\nsR} \omega \|\varphi_{k|t}\|^2  \Bigg).
\end{split}
\end{equation}
For each row i, the above objective function can be approximated to a convex optimization problem using a similar method as in Section \ref{sec:Convex Approximation}.

\subsubsection{Tracking Control: Expectation Objective Function}\label{sec:Expectation Objective Function MIMMO}
The expectation objective function for the MIMO system can be defined as
\begin{equation}\label{eq:cost function expected MIMO}
\begin{split}
    L_{\mathrm{exp}} \! = \! \min_{\{\mathrm{u}_{k|t}\}_{k=0}^\nsP} & \! \mathbb{E}\Bigg[ \sum_{k=0}^\nsP \Bigg( \sum_{i=1}^{n_\mathrm{y}} \left( \|z_{k+t}^{(i)} - \nu^{i^{\mathsf T}} \varphi_{k|t}\|^2 \right)
    -  \sum_{k=0}^{\nsR} \omega \|\varphi_{k|t}\|^2  \Bigg].
\end{split}
\end{equation}
Using the posterior distribution of the unknown parameter $\mu^i \sim \mathcal{N}(\tilde{{\mu}}_t^i, \tilde{\Sigma}_t^i)$, the above equation can be rewritten as
\begin{equation}
    \begin{split}
        L_{\mathrm{exp}}^{\mathrm{(ac)}}
        & \! = \!\! \min_{\{\mathrm{u}_{k|t}\}_{k=0}^\nsP} 
        \! \sum_{k=0}^\nsP  
        \! \Bigg( \! \sum_{i=1}^{n_{\mathrm{y}}} \Big( z_{k+t}^{(i)^2} \! + \! \varphi_{k|t}^{\mathsf T} \! \left( \! \tilde{\Sigma}_t^i \! + \! \tilde{\mu}_t^i \tilde{\mu}_t^{i^{\mathsf T}} \! \right) \! \varphi_{k|t}
        - 2z_{k+t}^{(i)} \tilde{\mu}_t^{i^\tr} \varphi_{k|t} \Big) \Bigg)
        - \sum_{k=0}^{\nsR} \omega \|\varphi_{k|t}\|^2 . 
    \end{split}
\end{equation}
The resulting problem can be solved using the same approach as in Section~\ref{sec:Expectation Objective Function}.

\section{Active Input Design for Informative Excitation}\label{sec:Actively Promoting Input Design}
 The optimization problems \eqref{eq:MPC worst-case} and \eqref{eq:MPC Expectation} find control inputs that balance exploitation and exploration, where the extent of exploration is regulated through the hyperparameter $\omega$. 
 By appropriately tuning $\omega$, one can encourage more exploratory behavior or place greater emphasis on tracking performance. However, one should note that the resulting information matrix
 \begin{equation}
     \mathrm{A}_0 + \sum_{t \geq 0} \varphi_t \varphi_t^\tr
 \end{equation}
may remain bounded, i.e., the largest eigenvalue of the posterior covariance matrix $\Sigma_t$ may not converge to zero. More precisely, while an exploration term is considered in the objective function, the resulting input sequence may not provide sufficient excitation for exploration.
To address this issue, we introduce a small random additive perturbation to the control input and show that, under suitable assumptions on the perturbation sequence, the resulting information matrix $\mxA_t$ grows sufficiently in all directions, ensuring that the maximum eigenvalue of $\Sigma_t$ goes to zero with high probability. This result provides a theoretical guarantee of asymptotic reduction of the parameter covariance while preserving the exploration-exploitation framework of the proposed controller.

Let the perturbation be denoted as $\vcv_t$, for $t \in \mathbb{Z}_+$.
Define the perturbed input as 
 \begin{equation}
 \begin{split}
     \hat{\mathrm{u}}_t & = \mathrm{u}_t + \mathrm{v}_t,\\
 \end{split}
 \end{equation}
 and the regressor as
\begin{equation}
 \begin{split}
     \hat{\varphi}_t &= \begin{bmatrix} 
            \hat{\mathrm{u}}_t^{\tr} &  
            \hat{\mathrm{u}}_{t-1}^{\tr} & \ldots & \hat{\mathrm{u}}_{t-n_\mathrm{g} + 1}^{\tr}
        \end{bmatrix}^{\tr}.
 \end{split}
 \end{equation}
We next show that, under suitable statistical assumptions on $\mathrm{v}_t$, the largest eigenvalue of $\Sigma_t$ converges to zero.

 \begin{theorem}\label{thm:convergence of A_t}
     Let $(\mathrm{v}_t)_{t=0}^\infty$ be an independent bounded zero-mean perturbation sequence added to the input, which satisfies the following conditions:
        \begin{equation}\label{eq:perturbation conditions}
        \begin{split}
            &\|\mathrm{v}_t\|_\infty \leq \bar{\mathrm{v}}, \\
            &\mathds{E}\left[\mathrm{v}_t \mathrm{v}_t^\tr\right] = \sigma_v^2 \mathbf{I},
        \end{split}
        \end{equation}
        where $\sigma_v$ is some small positive scalar. Then, we have
     \begin{equation}
         \lim_{t \to \infty} \lambda_{\max} (\tilde{\Sigma}_t) = 0,
     \end{equation}
     almost surely. Moreover, if $\mu \in \bigcap_{t = 0}^\infty \mathcal{C}_t$, then
     \begin{equation}
         \lim_{t \to \infty} \lambda_{\max} ({\Sigma}_t) = 0,
     \end{equation}
     almost surely.
 \end{theorem}
\begin{proof}
    The proof is provided in Appendix~\ref{app: proof of convergence of A_t}.
\end{proof}

\begin{remark}
\normalfont
    In practice, there exist many random signals that satisfy the above statistical properties. For example, $\mathrm{v}_t$ may be chosen to follow a uniform distribution in the interval $[-a, a]$. With a suitable choice of $a$, such a signal satisfies the required statistical properties.
\end{remark}

\begin{figure*}[t!]
   \centering
   \includegraphics[trim={0cm 0cm 0cm 0cm},clip]{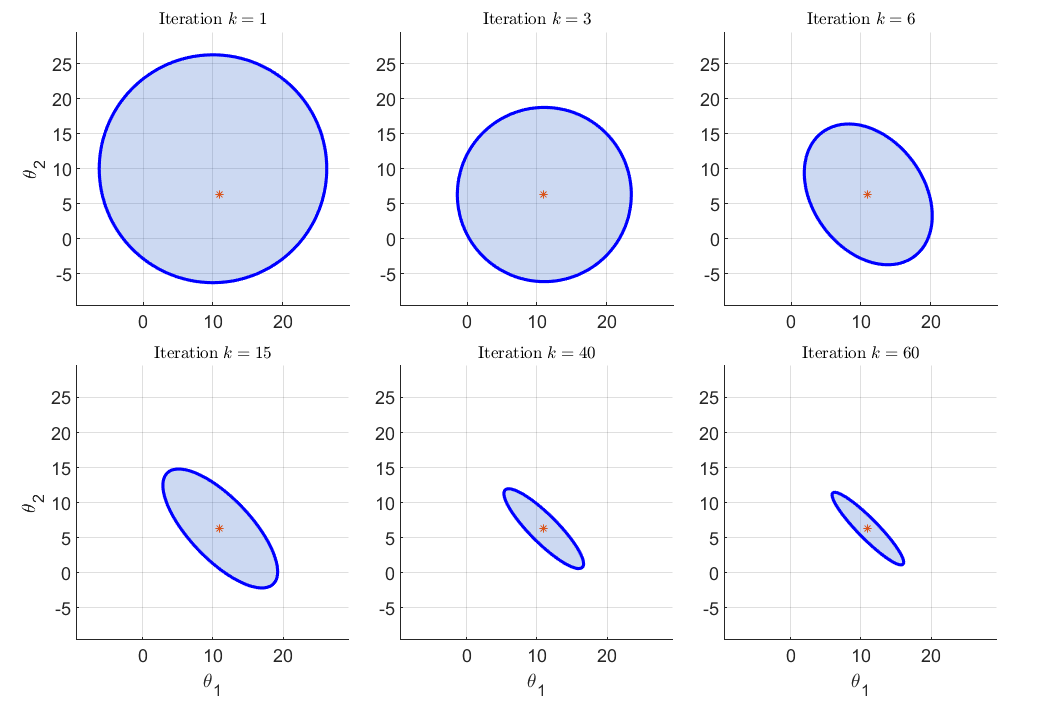}
   \caption{Evolution of the robust parameter set over time $t$. The marker represents the true parameter.}
   \label{img:evolution of parameter set}
\end{figure*}

Theorem~\ref{thm:convergence of A_t} shows that, with an arbitrarily small perturbation satisfying conditions \eqref{eq:perturbation conditions}, the maximal eigenvalue of the parameter covariance matrix $\tilde{\Sigma}_t$ converges to zero almost surely. The same convergence result holds for $\Sigma_t$ under the additional condition $\mu \in \bigcap_{t=0}^\infty \mathcal{C}_t$, which has a probability at least $1-\delta$. Consequently, we can choose a sufficiently small bound $\bar{\mathrm{v}}$ such that
    \begin{equation}
        |\nu^\tr \vcv_t| \in \mathcal{E}_v
    \end{equation}
holds for all $\nu \in \mathcal{C}_0$, where $\mathcal{E}_v$ is a given compact set.

Moreover, define $\mathcal{\bar{V}} := \{\vcv \in \mathbb{R}^{n_\mathrm{u}} | \| \vcv\|_\infty \leq \bar{\mathrm{v}}\}$.
The optimization program for active input design is as follows, 
\begin{subequations}\label{eq:optimization active input design}
\begin{align}
    &\min_{\{\mathrm{u}_{k|t}\}_{k=0}^{\nsP}} \; \, J(\mathbf{u}_{t,\nsP}) - \omega I(\mathbf{u}_{t,\nsR}) \\
    & \quad \, 
    \text{s.t.} 
    \quad \quad
    \mathrm{u}_{k|t} \in \mathcal{U} \ominus \mathcal{\bar{V}}, \\
    & \qquad \qquad \; \,  
    \varphi_{\nsP|t} \in \mathcal{R}_t \ominus \mathcal{\bar{V}}^{n_g},\\
    & \qquad \qquad \; \,  
    \mathds{P} \left( {\mathrm{y}}_{k|t} \in \mathcal{Y}_{\mathrm{safe}} \ominus \mathcal{E}_v \right) \geq 1- \delta_{\tau_t}.
\end{align}
\end{subequations}
After solving the optimization problem, $\mathrm{u}_{t}$ is obtained from the optimal solution. The perturbed input $\hat{\mathrm{u}}_t$ is then applied to the system to collect data.
\begin{corollary}
    The optimization problem \eqref{eq:optimization active input design} is recursively feasible, and the resulting closed-loop system satisfies the safety constraints with probability at least $1-\delta$.
\end{corollary}
\begin{proof}
    The proof of safety and recursive feasibility is similar to that in Theorem~\ref{thm:safety} and Theorem~\ref{thm:recursive feasibility}, respectively.
\end{proof}

\section{Numerical Experiments}\label{sec: Numerical Experiments}
In this section, we present several numerical experiments to evaluate the proposed Safe Performance-aware Online Input Design (SPOID) framework.

\subsection{Two-Dimensional Impulse Response Case}\label{sec: Two-Dimensional Impulse Response}

The first numerical experiment evaluates parameter estimation performance. In this experiment, the truncation order of the impulse response $n_{\mathrm{g}}$ is selected to be 2 for the sake of visualization. The initial mean and covariance matrix of the real parameter is set as ${\mu}_0 = \begin{bmatrix} 10 \\ 10 \end{bmatrix}$ and ${\Sigma}_0 = \begin{bmatrix} 6 & 0  \\ 0 & 6 \end{bmatrix}$. The true parameter is randomly sampled from this Gaussian distribution. The tracking reference $z_t$ is designed using a piecewise constant function. Algorithm \ref{Al: SID} with the worst-case cost function is applied using $\nsP=4$ and $\nsR=2$. The simulation time steps is 60.

Figure \ref{img:evolution of parameter set} illustrates the evolution of the robust parameter set throughout the identification process. At the initial stage, the robust parameter set is relatively large, reflecting significant uncertainty in the model parameters. To reduce this uncertainty, the algorithm emphasizes exploration during early iterations, actively collecting informative data. As a consequence, the size of the uncertainty set gradually decreases. As the algorithm progresses and more data become available, the parameter estimates become increasingly accurate. The need for exploration diminishes, and the algorithm focuses more on exploitation using the refined parameter estimates. Accordingly, the ellipsoidal uncertainty set continues to contract at a slower rate.

\begin{figure}[t!]
   \centering
   \includegraphics[trim={0cm 0cm 0cm 0cm},clip]{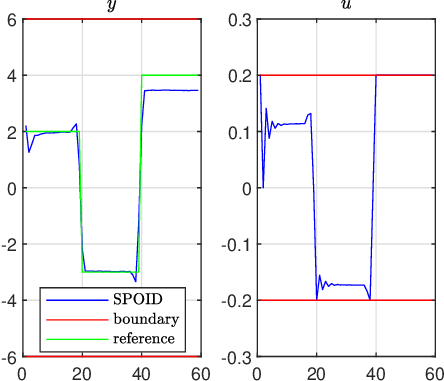}
   \caption{Evolution of constrained tracking performance under parameter uncertainty.}
   \label{img:tracking}
\end{figure}

This behavior is also illustrated in Figure~\ref{img:tracking}. In the initial stage, the output trajectory generated by SPOID deviates from the reference signal due to the emphasis on exploration. After several time steps, as the parameter estimates improve, the algorithm gradually shifts toward exploitation, resulting in an output trajectory that more closely tracks the reference. When the reference changes, a temporary increase in tracking error is observed. This behavior may be attributed to the finite prediction horizon and the trade-off between exploration and exploitation.

\subsection{High-Dimensional Impulse Response}\label{sec: High-Dimensional Impulse Response}
In this experiment, we evaluate the performance of the proposed algorithm using a system with a larger truncation parameter. The true parameter corresponds to a randomly generated system with a decaying impulse response. The truncation parameter $n_{\mathrm{g}}$ is selected as the minimal integer subject to 
\begin{equation}
    \frac{\sum_{i\geq n_{\mathrm{g}}} \mathrm{g}_i^2}{\sum_{j \geq 0} \mathrm{g}_j^2} \leq 4e^{-4}.
\end{equation}
The above equation indicates that the energy of the truncated tail is small relative to the total impulse response energy.
The prior mean ${\mu}_0$ is a one vector and ${\Sigma}_0$ is $8^2 \mathbf{I}$. The total number of simulation steps is $T = 200$. The tracking reference is designed as $\mathrm{z}_t = 240 \sin (4.5\pi t/T)$. The safety constraints for the input and output are $|\mathrm{u}_t| \leq 5$ and $|\mathrm{y}_t| \leq 200$, respectively. Thus, the reference can exceed the output safety bounds during certain periods. The proposed Algorithm \ref{Al: SID} is applied to the system with the worst-case and expectation objective functions. The horizon $\nsP$ is set at 30 and $\nsR$ is set as 5. Both methods are evaluated over 200 Monte Carlo realizations with independently generated Gaussian noise sequences.

\begin{figure}[t!]
   \centering
   \includegraphics[trim={0cm 0cm 0cm 0cm},clip]{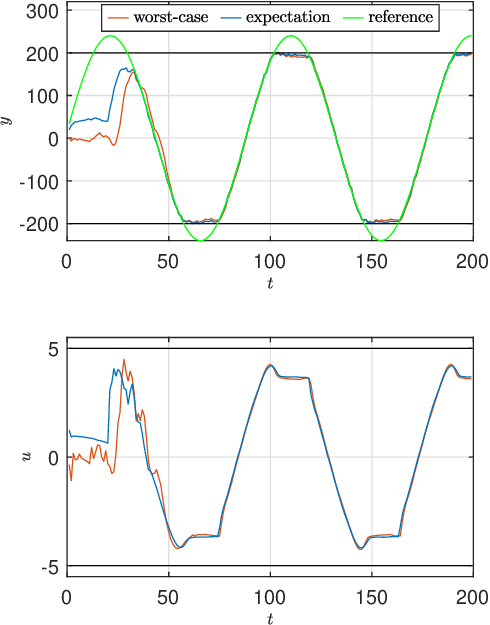}
   \caption{Comparison of worst-Case and expectation identification for constrained tracking}
   \label{img:tracking_large}
\end{figure}

Figure \ref{img:tracking_large} illustrates the tracking performance of SPOID under the two cost formulations for a single realization. Both formulations have similar overall tracking behavior. Initially, the parameter uncertainty is large. Consequently, both methods prioritize exploration over tracking performance, and the system output does not closely follow the reference trajectory. Since the matrix $\varphi_k \varphi_k^{\tr}$ is rank one, the excitation introduced at each time step affects only a single direction in the parameter space. Therefore, each time step contributes excitation along at most one direction in the parameter space. As excitation accumulates across different directions in the parameter space, the robust parameter set contracts significantly. As the uncertainty set shrinks, the controller becomes less conservative, allowing the system output to track the reference more closely. As learning continues, the robust parameter set further contracts, which enhances tracking performance. However, around time step 60, the reference trajectory approaches and subsequently exceeds the output constraint boundary. To satisfy the probabilistic safety constraint, the controller limits the output response as the reference exceeds the admissible region. Consequently, the output remains within the prescribed safety bounds in the illustrated realization until the reference moves away from the boundary.
Despite this, the two formulations also have distinct control behaviors.
In this experiment, the worst-case formulation exhibits more conservative control behavior than the expectation-based formulation. Consequently, the expectation-based method achieves faster excitation during the initial stage, allowing its output to approach the reference trajectory more rapidly.


\begin{figure}[t!]
   \centering   
   \includegraphics[trim={0cm 0cm 0cm 0cm},clip]{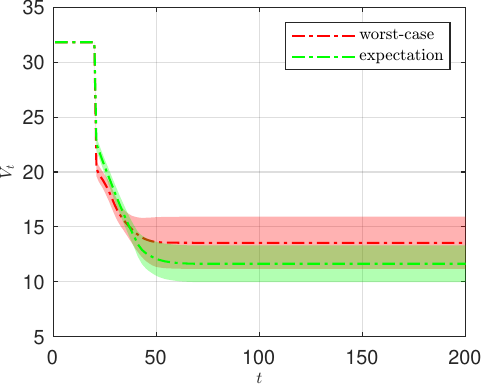}
   \caption{Evolution of the log-volume for the robust parameter set. The dashed line represents the mean volume over 200 Monte Carlo realizations, while the shaded region corresponds to the interval between the mean minus one standard deviation and the mean plus one standard deviation. }
   \label{img:volumn}
\end{figure}

\begin{figure*}[t!]
   \centering
   \includegraphics[trim={0cm 0cm 0cm 0cm},clip]{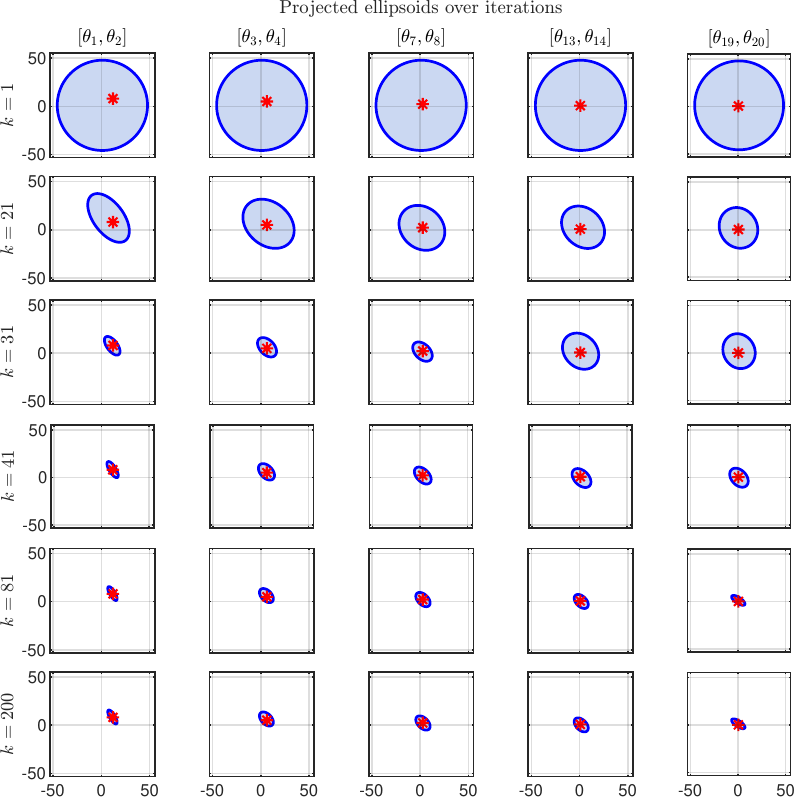}
   \caption{Evolution of the projected robust parameter set over time $t$. The marker represents the true parameter.}
   \label{img:parameter_covergence}
\end{figure*}

Figure \ref{img:volumn} shows the evolution of the robust parameter set. For visualization, we use the log-volume of the ellipsoid, defined as
\begin{equation}
    V_t = \log \mathrm{vol}(\mathcal{{C}}_t).
\end{equation}
In this experiment, a noticeable contraction of the robust parameter set occurs after the number of collected observations exceeds the parameter dimension. This behavior suggests that the accumulated data provide sufficient excitation across the parameter directions, leading to a substantial reduction in uncertainty.
Moreover, the robust parameter set contracts more slowly under the worst-case cost formulation than under the expectation-based formulation, which is consistent with the results observed in Figure~\ref{img:tracking_large}.


We also examine the evolution of parameter estimates over time. Similar trends are observed for Algorithm~\ref{Al: SID} under both the worst-case and the expectation cost formulation. For brevity, we present only the results for the expectation cost formulation as shown in Figure~\ref{img:parameter_covergence}. For visualization, the confidence ellipsoid is projected onto five two-dimensional subspaces.
The figure shows that the uncertainty of the parameters decreases progressively as time increases. In particular, the uncertainty associated with lower-index impulse response coefficients shrinks earlier and at a faster rate. 
This behavior may be related to the decaying impulse response of the system, as coefficients with larger magnitudes have a stronger influence on the predicted output and may therefore receive more effective excitation through the performance-aware input design.
\section{Conclusion}\label{sec:Conclusion}
In this paper, we propose a framework for the safe performance-aware online identification of an unknown system without requiring informative initial data. Safety constraints are incorporated to provide high-probability guarantees of safe operation throughout the online input design procedure. A Bayesian method is employed to update the parameter estimates online. The objective function is designed to balance exploitation and exploration: the exploitation term is formulated using either a worst-case or an expected cost, while the exploration term promotes active learning. To facilitate online implementation, we derive a convex optimization formulation for the expectation-based cost and develop a convex approximation for the worst-case cost. The effectiveness of the proposed approximation is evaluated through numerical experiments. The algorithm is then extended to MIMO systems. We also introduce an active input design strategy to ensure asymptotic convergence of the parameter covariance to zero under appropriate conditions. Finally, numerical experiments are conducted to assess the proposed framework in terms of control performance and uncertainty reduction.
\appendix
\section{Auxiliary Lemma}    
\begin{lemma}\label{lem:|c^T a|}
    Assume $\mathrm{c} \in \mathbb{R}^d$ is a non-zero vector and $\mathrm{a} \in \mathbb{R}^d$ is in an ellipsoid such that $(\mathrm{a} - \bar{\mathrm{a}})^\tr \mathrm{S} (\mathrm{a} - \bar{\mathrm{a}}) \leq \beta$, where $\mathrm{S}$ is a positive definite matrix. The following inequality holds
    \begin{equation}
        |\mathrm{c}^\tr \mathrm{a}| \leq \left( \beta \mathrm{c}^\tr\mathrm{S}^{-1} \mathrm{c} \right)^{\frac{1}{2}} + |\bar{\mathrm{a}}^\tr \mathrm{c}|.
    \end{equation}
\end{lemma}
\begin{proof}
    Define $\mathrm{S} := \mathrm{Q}\mathrm{Q}^\tr$ and $\mathrm{b} := \mathrm{Q}^\tr(\mathrm{a} - \bar{\mathrm{a}})$. Thus, we have $\mathrm{b}^\tr\mathrm{b} \leq \beta$.
    Consider $\max \mathrm{c}^\tr \mathrm{a}$, which can be written as
    \begin{equation}\label{eq:max c^T a}
        \begin{split}
            \max_{\|\mathrm{b}\|_2^2 \leq \beta} \mathrm{c}^\tr \mathrm{a}
            =& \max_{\|\mathrm{b}\|_2^2 \leq \beta} \mathrm{c}^\tr (\mathrm{Q}^{- \tr} \mathrm{b} + \bar{\mathrm{a}})\\
            =& \left( \beta \mathrm{c}^\tr\mathrm{S}^{-1} \mathrm{c} \right)^\frac{1}{2} + \bar{\mathrm{a}}^\tr \mathrm{c}.
        \end{split}
    \end{equation}
    Similarly, we have
    \begin{equation}\label{eq:min c^T a}
        \begin{split}
            \min_{\|\mathrm{b}\|_2^2 \leq \beta} \mathrm{c}^\tr \mathrm{a}
            =& -\left( \beta \mathrm{c}^\tr\mathrm{S}^{-1} \mathrm{c} \right)^{\frac{1}{2}} + \bar{\mathrm{a}}^\tr \mathrm{c}.
        \end{split}
    \end{equation}
    Combining \eqref{eq:max c^T a} and \eqref{eq:min c^T a} concludes the proof.
\end{proof}

\begin{lemma}\label{lem:Strong Law of Large Numbers}\cite{hall2014martingale}
    Let $\{X_n, \mathcal{F}_n\}_{n\geq 1}$ be a martingale difference sequence  defined on a probability space almost surely. Let $S_n = \sum_{k=1}^n X_k$. 
    If 
    \begin{equation}
        \sup_{n \geq 1} \mathds{E}[\|X_n\|^2] < \infty,    
    \end{equation}
    is satisfied.
    Then, we have
    \begin{equation}
        \lim_{n \to \infty} \frac{S_n}{n} = 0,
    \end{equation}
    almost surely.
\end{lemma}

\section{Proof of Theorem~\ref{thm:recursive feasibility}}\label{app:proof of recursive feasibility}
    We prove recursive feasibility for \eqref{eq:MPC worst-case} by induction. The proof for \eqref{eq:MPC Expectation} follows analogously. Assume that the optimization problem is feasible at time step $k$. We need to show that it is feasible at time step $k+1$. 
    
    Let $\lambda_t^*, c_t^*, \{\mathrm{u}_{k|t}^*, \bar{\lambda}_{k|t}^*, \mathrm{D}_{k|t}^*, \bar{\mathrm{D}}_{k|t}^*\}_{k=0}^\nsP$ be the solution of \eqref{eq:MPC worst-case} at time step $t$. Define
    \begin{equation}\label{eq:backup solution 1}
    \begin{split}
        \mathrm{u}_{k|t+1} = 
        \begin{cases}
            \mathrm{u}_{k+1|t}^*, & k \in \mathbf{Z}_{\nsP-1},\\
            \mathrm{K} \varphi_{\nsP|t}^*, & k=\nsP,
        \end{cases} \\
    \end{split}
    \end{equation}
    where $\mathrm{K}$ is pre-computed before \eqref{eq:terminal constraints}. 

     For constraints \eqref{eq:MPC worst-case 2}-\eqref{eq:MPC worst-case 4}, consider the following values. Define
    \begin{equation}\label{eq:backup solution 2}
    \begin{split}
        \mathrm{\lambda}_{t+1} &= \mathrm{\lambda}_{t}^*, \\
        {\mathrm{D}}_{k|t+1} & = 
        \begin{cases}
            {\mathrm{D}}_{k+1|t}^*, & k \in \mathbf{Z}_{\nsP-2}, \\
            \varphi_{k+1|t+1} \varphi_{k+1|t+1}^\tr + \mathrm{D}_{k+1|t+1}, & k = \nsP\!-\!1,\\
            \mathbf{I}, & k = \nsP.
        \end{cases}\\
        c_{t+1} &= \mathrm{v}^\tr\mathrm{v} - p,
    \end{split}
    \end{equation}
    where
    \begin{equation}
        \begin{split}
            p &= \lambda_{t+1} a_{t+1}-\! \lambda \beta_{t+1}  \!\!-\! \sum_{k=0}^\nsP\! z_{k+t+1}^2,\\
            \mathrm{v} &= - \lambda_{t+1} d_{t+1} + \sum_{k=0}^\nsP z_{k+t+1}\varphi_{k|t+1}.
        \end{split}
    \end{equation}
    We show that the proposed $c_{t+1}, \lambda_{t+1}, \{\mathrm{u}_{k|t}, \mathrm{D}_{k|t}\}_{k=0}^\nsP$ satisfy constraints \eqref{eq:MPC worst-case 2}-\eqref{eq:MPC worst-case 4}.
    
    First, consider the constraint \eqref{eq:MPC worst-case 2} at time step $k+1$,
    \begin{equation}
        \begin{bmatrix}
            \mathrm{D}_{\nsP|t+1} & - \lambda_{t+1} d_{t+1} + \sum_{k=0}^\nsP z_{k+t+1}\varphi_{k|t+1} \\
        \star & \lambda_{t+1} a_{t+1} \!\!-\! \lambda \beta_{t+1} \!+ c_{t+1} \!\!-\! \sum_{k=0}^\nsP\! z_{k+t+1}^2
        \end{bmatrix}
        \!\succeq\! 0.
    \end{equation}
    Let $\lambda_{t+1}$ and $\mathrm{u}_{k|t+1}$ be constructed as in \eqref{eq:backup solution 1} and \eqref{eq:backup solution 2}, respectively. Using Schur complement, the above equation is equivalent to 
    \begin{equation}
        c_{t+1} + p - \mathrm{v}^\tr \mathrm{D}_{\nsP|t+1}^{-1} \mathrm{v} \geq 0, 
    \end{equation}
    One can find for any given $\mathrm{D}_{\nsP|t+1}^{-1}$, the above constraint holds by choosing sufficiently large $c_{t+1}$. For example, in our  construction \eqref{eq:backup solution 2}, we use $\mathrm{D}_{\nsP|t+1} = \mathbf{I}$ and $c_{t+1} = \mathrm{v}^\tr\mathrm{v} - p$ as a feasible solution.
    
    For constraint \eqref{eq:MPC worst-case 3}, one can notice since $\mathrm{A}_{t+1} \succeq \mathrm{A}_t$, from Schur complement, we have 
    \begin{equation}
        \lambda \mathrm{A}_{t+1} \succeq \lambda \mathrm{A}_{t} \succeq \mathrm{D}_{0|t}^*,
    \end{equation}
    which implies
    \begin{equation}
        \lambda \mathrm{A}_{t+1} - \mathrm{D}_{0|t+1} = \lambda \mathrm{A}_{t+1} - \mathrm{D}_{1|t}^* \succeq \mathrm{D}_{0|t}^* - \mathrm{D}_{1|t}^*.  
    \end{equation}
    From the assumption, we have
    \begin{equation}
    \begin{bmatrix}
        \mathrm{D}_{0|t}^* - \mathrm{D}_{1|t}^* & \varphi_{1|t}^*\\
        \star & 1
    \end{bmatrix}
    \succeq 0,
    \end{equation}
    which implies that $\eqref{eq:MPC worst-case 3}$ holds.

    Next, for $k_1 \in \mathbf{Z}_{\nsP-2}$, \eqref{eq:MPC worst-case 4} is directly satisfied based on the construction. For $k_1 = \nsP-1$, since 
    \begin{equation}
        {\mathrm{D}}_{\nsP-1|t+1} = \varphi_{\nsP|t+1} \varphi_{\nsP|t+1}^\tr + \mathrm{D}_{\nsP|t+1},
    \end{equation}
    which implies that
    \begin{equation}
        \begin{bmatrix}
        \mathrm{D}_{\nsP-1|t+1} - \mathrm{D}_{\nsP|t+1} & \varphi_{\nsP|t+1}\\
        \star & 1
        \end{bmatrix}
        \succeq 0.
    \end{equation}
    Thus, \eqref{eq:MPC worst-case 4} holds for $k_1 \in \mathbf{Z}_{\nsP-1}$.

     The next three constraints \eqref{eq:MPC worst-case 5}-\eqref{eq:MPC worst-case 7} can be shown by the construction of robust parameter set and online terminal set. For $k_2 \in \mathbf{Z}_{\nsP-1}$, \eqref{eq:MPC worst-case 5}-\eqref{eq:MPC worst-case 6} hold from the construction of robust parameter set. Recall as mentioned in Section \ref{sec:Robust Safety Constraints}, \eqref{eq:MPC worst-case 5} and \eqref{eq:MPC worst-case 6} are equivalent to
     \begin{equation}
        \|\nu^\tr \varphi_{k_2|t+1}\| \leq \bar{\mathrm{y}}, \quad \forall \nu \in \mathcal{C}_{t+1}, \quad k_2 \in \mathbf{Z}_{\nsP}.
    \end{equation}
    From assumption, we have
    \begin{equation}
        \|\nu^\tr \varphi_{k_2|t}^*\| \leq \bar{\mathrm{y}}, \quad \forall \nu \in \mathcal{C}_{t}, \quad k_2 \in \mathbf{Z}_{\nsP}.
    \end{equation}
    Since $\varphi_{k_2|t+1} = \varphi_{k_2+1|t}^*$ and $\mathcal{C}_{t+1} \subseteq {\mathcal{C}_t}$, \eqref{eq:MPC worst-case 5} and \eqref{eq:MPC worst-case 6} hold for $k_2\in \mathbf{Z}_{\nsP-1}$. 
    
    For $k_2=\nsP$, we use the property from terminal set. Since $\mathrm{u}_{\nsP|t+1}  = \mathrm{K}\varphi_{\nsP|t}^* = \mathrm{K}\varphi_{\nsP-1|t+1}$ and $\varphi_{\nsP-1|t+1} = \varphi_{\nsP|t}^* \in \mathcal{R}_{t} \in \mathcal{R}_{t+1}$, from \eqref{eq:terminal constraints 3}, we have
    \begin{equation}
    \begin{split}
        |\nu^\tr \varphi_{\nsP|t+1}| &\leq \bar{y}, \quad \forall \nu \in \mathcal{C}_{t+1},\\
    \end{split}
    \end{equation}
    which implies there exists $\bar{\lambda}_{k_2|t} \geq 0$ and $\bar{\mathrm{D}}_{k_2|t}$ such that \eqref{eq:MPC worst-case 5} and \eqref{eq:MPC worst-case 6} hold.
    
    For the last constraint, from \eqref{eq:terminal constraints 1}, we have
    \begin{equation}
        \varphi_{\nsP|t+1}^\tr\mathrm{S}_t\varphi_{\nsP|t+1} \leq \varphi_{\nsP|t}^\tr\mathrm{S}_t\varphi_{\nsP|t} \leq 1,
    \end{equation}
    which implies that $\varphi_{\nsP|t+1} \in \mathcal{R}_t \in \mathcal{R}_{t+1}$ and concludes the proof.
    \qed

\section{Proof of Theorem~\ref{thm:convergence of A_t}}\label{app: proof of convergence of A_t}
    Define vector $\zeta_t \in \mathbb{R}^{n_g n_{\mathrm{u}}}$ as
    \begin{equation}
        \zeta_t = \begin{bmatrix}
            \mathrm{v}_t \\ \mathrm{v}_{t-1} \\ \vdots \\ \mathrm{v}_{t-n_g+1}
        \end{bmatrix}.
    \end{equation}
    Under this notation, we have
    \begin{equation}
        \hat{\varphi}_t = \varphi_t + \zeta_t.
    \end{equation}
    Let $k_j = n_g j$, $j \in \mathbb{N}$. Since the truncation length is $n_g$, the vectors $\zeta_{k_j}$ are mutually independent. Moreover, we have
    \begin{equation}
    \begin{split}
        &\mathds{E}[\zeta_{k_j}] = 0,\\
        &\mathds{E}[\zeta_{k_j}\zeta_{k_{j-1}+1}^\tr] \succeq \sigma_v^2 \mathbf{I}.
    \end{split}   
    \end{equation}
    Define $Y_{k_j}$ as
    \begin{equation}
    \begin{split}
        Y_{j} &:= \hat{\varphi}_{k_j}\hat{\varphi}_{k_j}^\tr - \mathds{E}\left[ \hat{\varphi}_{k_j}\hat{\varphi}_{k_j}^\tr | \mathcal{F}_{k_{j-1}+1}\right],\\
        Z_{m} &:= \sum_{j=1}^m Y_{j}.
    \end{split}
    \end{equation}
    Note since $\hat{\varphi}_t$ is applied to the system, the history is redefined as
    \begin{equation} 
        \mathcal{F}_t := \sigma\big(\{\hat{\varphi}_0, y_0, \ldots, \hat{\varphi}_{t-1}, y_{t-1}\} \big).
    \end{equation}
    Then 
    \begin{equation}
        \mathds{E}\left[Y_{j} | \mathcal{F}_{k_{j-1}+1} \right] = 0.
    \end{equation}
    which implies that $Z_m$ is a matrix martingale. Since $\varphi_{k_j}$ and $\zeta_{k_j}$ are both bounded, using 
    Lemma~\ref{lem:Strong Law of Large Numbers}, we have
    \begin{equation}\label{eq:probability of sum Z_m}
        \mathds{P} \left( \lim_{m \to \infty} \frac{1}{m} Z_m = 0 \right) = 1.
    \end{equation}
    Define
    \begin{equation}
        \begin{split}
            \mxM_j &= \mathds{E}[\zeta_{k_j} \hat{\varphi}_{k_j}^\tr | \mathcal{F}_{k_{j-1}+1}],
            \\
            \mxD_j &= \mathds{E}[\zeta_{k_j} {\zeta}_{k_j}^\tr | \mathcal{F}_{k_{j-1}+1}].
        \end{split}
    \end{equation}
    Since $\vcv_t$ is mutually independent and also independent of ${\vcu}_{s}$, for any $s\leq t$, $\mxM_j$ is upper triangular satisfying
    \begin{equation}
        [\mxM_j]_{ii} = \sigma_v^2,
    \end{equation}
    for any $i \in \mathbb{N}_{\Nu n_\mathrm{g}}$.
    Since the inputs and perturbations are uniformly upper bounded, one can see that $\|\mxM_j\|_2$ is also upper bounded. Thus, there exists deterministic constants $C_1, C_2 >0$, such that
    \begin{equation}
    \begin{split}
        \mxM_j^\tr \mxM_j &\succeq C_1 \mathbf{I},\\
        \mxD_j &\preceq C_2\mathbf{I}.
    \end{split}
    \end{equation}
    Using Schur complement, we have
    \begin{equation}
    \begin{split}
        \mathds{E}\left[\hat{\varphi}_{k_j}\hat{\varphi}_{k_j}^\tr | \mathcal{F}_{k_{j-1}+1}\right] 
        &\succeq \mxM_j^\tr \mxD_j^{-1} \mxM_j
        \\
        &\succeq \frac{C_1}{C_2} \mathbf{I}.
    \end{split}
    \end{equation}
    Therefore, one can see that
    \begin{equation}
        \begin{split}
            \lim_{m \to \infty}\frac{1}{m} \sum_{j=1}^m \hat{\varphi}_{k_j}\hat{\varphi}_{k_j}^\tr
            =& \lim_{m \to \infty}\frac{1}{m} \sum_{j=1}^m \mathds{E}\left[ \hat{\varphi}_{k_j}\hat{\varphi}_{k_j}^\tr | \mathcal{F}_{k_{j-1}+1} \right] + \frac{1}{m} Z_m,\\
            \succeq& \lim_{m \to \infty} \frac{C_1}{C_2} \mathbf{I} + \frac{1}{m} Z_m.
        \end{split}
    \end{equation}
    Applying \eqref{eq:probability of sum Z_m}, we have
    \begin{equation}
        \mathds{P} \left(\liminf_{m \to \infty} \lambda_{\min} \left( \frac{1}{m} \sum_{j=1}^m \hat{\varphi}_{k_j}\hat{\varphi}_{k_j}^\tr \right) \geq \frac{C_1}{C_2} \right) = 1.
    \end{equation}
    Moreover, define $K:= \lfloor \frac{T}{n_g}\rfloor$, we have
    \begin{equation}
        \begin{split}
            \tilde{\mxA}_T \succeq \tilde{\mxA}_0 + \sum_{j=1}^{K} \frac{1}{\sigma_w^2} \hat{\varphi}_{k_j}\hat{\varphi}_{k_j}^\tr.
        \end{split}
    \end{equation}
    Combining the above two equations, we have
    \begin{equation}
        \mathds{P} \left(\liminf_{T \to \infty} \lambda_{\min} \left( \frac{1}{K} (\tilde{\mxA}_T - \tilde{\mxA}_0) \right)\succeq \frac{C_1}{C_2 \sigma_w^2}  \right) = 1.
    \end{equation}
    Since $K= \lfloor \frac{T}{n_g}\rfloor$, the above equation implies $ \lambda_{\max} (\tilde{\mxA}_t^{-1}) = O(1/t)$. Thus, we have
    \begin{equation}
        \mathds{P} \left( \lim_{T \to \infty} \lambda_{\max} \left(\tilde{\mxA}_T^{-1}\right) = 0  \right) = 1,
    \end{equation}
    or equivalently
    \begin{equation} \label{eq:lim lambda A_tilde}
        \mathds{P} \left( \lim_{T \to \infty} \tilde{\Sigma}_T = 0  \right) = 1.
    \end{equation}
    Next, we show that $\mxA_t^{-1}$ also converges to zero on the event of $\mu \in \bigcap_{\tau=0}^\infty \mathcal{C}_\tau$, which has probability greater than $1-\delta$ by Theorem~\ref{thm:probability constraint}. 
    From the update law in \eqref{eq:robust parameter set update}, it is sufficient to show that there does not exist $i$, such that $\mathcal{C}_t = {\mathcal{C}}_{t_i}$, for all $t \geq t_i$, where $t_i$ is a certain time that $[\tilde{\beta}_{t_i}^{\min}, \tilde{\beta}_{t_i}^{\max}]$ is not an empty set. We show it by contradiction. If such $i$ exists, then the update indices are countable. Since the boundary of each ellipsoid has posterior probability zero, on the event $\mu \in \bigcap_{\tau=0}^\infty \mathcal{C}_\tau$, we also have $\mu \in \mathrm{int}(\mathcal{C}_{t_i})$ for every update time $t_i$, almost surely.
    Then there exists $r_i >0$ such that
    \begin{equation}
            \mathcal{B}(\mu, r_i) \subseteq \mathcal{C}_{t_i},
    \end{equation}
    where $\mathcal{B}(\mu, r_i)$ is the ball centered at $\mu$ and has radius $r_i$.
    Consider difference between $\tilde{\mu}_t$ and $\mu$ as
    \begin{equation}
        \begin{split}
            \tilde{\mu}_t  - \mu &= \tilde{\mxA}_t^{-1} \left( \mxA_0(\mu_0 - \mu)  + \frac{1}{\sigma_w^2} \sum_{k=1}^t \hat{\varphi}_k \vcw_k \right).
        \end{split}
    \end{equation}
    Define $L_t := \sum_{k=1}^t \hat{\varphi}_k \vcw_k$. Since $\varphi_k$ is bounded and $\vcw_k$ is Gaussian with finite variance and independent of $\varphi_k$, one can see that $L_t$ is a martingale. Using Lemma~\ref{lem:Strong Law of Large Numbers} again, we have
    \begin{equation}
        \lim_{t \to \infty} \frac{1}{t} L_t = 0.
    \end{equation}
    Since $\|\mu\|, \|\mu_0\| < \infty$ and $\lambda_{\max} (\tilde{A}_t^{-1}) = O(1/t)$, we have
    \begin{equation}
        \lim_{t \to \infty} \|\tilde{\mu}_t - \mu\| = 0. 
    \end{equation}
    Thus, for fixed $q_{i} := \chi^2_{n_{\mathrm{g}}}(1 - \delta_{i})$, there exists $t \geq t_i$, such that
    \begin{equation}
        \sqrt{\frac{q_{i+1}}{\lambda_{\min}(\tilde{\mxA}_{t})}} + \|\tilde{\mu}_t - \mu\| < r_i,
    \end{equation}
    which implies $\tilde{\mathcal{C}}_t(\tilde{\beta}^{\min}_t) \subsetneq \mathcal{B}(\mu, r_i) \subseteq \mathcal{C}_{t_i}$.
    Thus, $[\tilde{\beta}_{t_i}^{\min}, \tilde{\beta}_{t_i}^{\max}]$ is not an empty set. This implies that for any $i$, there exists $t \geq t_i$, such that $\mathcal{C}_t \subsetneq \mathcal{C}_{t_i}$, which leads to a contradiction. Therefore, for any $i$, there exists a time $t_i$, such that $\mathcal{C}_{t_i}$ updates to $\tilde{\mathcal{C}}_{t_t}(\tilde{\beta}_{t_i}^{\max})$ and $\mxA_{t_i} = \tilde{\mxA}_{t_i}$. From \eqref{eq:lim lambda A_tilde}, we have
    \begin{equation} 
        \mathds{P} \left( \lim_{T \to \infty} {\Sigma}_T = 0  \Bigm\vert \mu \in \bigcap_{\tau=0}^\infty \mathcal{C}_\tau \right) = 1,
    \end{equation}   
    This concludes the proof. \qed
\bibliographystyle{IEEEtran}
\bibliography{mainbib}



\end{document}